\magnification = 1150
\documentstyle{amsppt}
\define\>{\rangle}
\define\<{\langle}
\define\st{\,|\,}

\define\Lo{\Bbb L}
\define\R{\Bbb R^}
\define\Sph{\Bbb S^}
\define\Min{\Bbb L^}
\define\Minn{\Min2_{\alpha,\omega}}
\define\dn{\delta^{t_n}_{\rho_n,x_n}}

\define\({\left(}
\define\){\right)}

\define\fcb{\hat\partial}
\define\Del{\nabla}
\define\dln{\delta^{t_n}_{\rho_n,x_n}}
\define\AdS{\Bbb A\text d\Bbb S}
\define\Sch{\Bbb S\text{ch}}
\define\Int{_{\text{int}}}
\define\ext{_{\text{ext}}}
\define\rmin{r_{\text{min}}}
\define\rmax{r_{\text{max}}}
\define\reh{r_{\text{EH}}}
\define\eh{_{\text{EH}}}

\input amstex

\topmatter
 
\title Complete Affine Connection in the Causal Boundary: \\ Static, Spherically Symmetric Spacetimes
\endtitle

\rightheadtext
{Complete Connection on Boundary}

\author
Steven (Stacey) G. Harris
\endauthor

\address
Department of Mathematics, Saint Louis University,
St. Louis, MO 63103, USA \endaddress
\email
harrissg\@slu.edu \endemail

\abstract
The boundary at $\Cal I^+$, future null infinity, for a standard static, spherically symmetric spactime is examined for possible linear connections.  Two independent methods are employed, one for treating $\Cal I^+$ as the future causal boundary, and one for treating it as a conformal boundary (the latter is subsumed in the former, which is of greater generality).  Both methods provide the same result: a constellation of various possible connections, depending on an arbitrary choice of a certain function, a sort of gauge freedom in obtaining a natural connection on $\Cal I^+$; choosing that function to be constant (for instance) results in a complete connection.  

Treating $\Cal I^+$ as  part of the future causal boundary, the method is to impute  affine connections on null hypersurfaces going out to $\Cal I^+$, in terms of a transverse vector field on each null hypersurface (there is much gauge freedom on choice of the transverse vector fields).  Treating $\Cal I^+$ as part of a conformal boundary, the method is to make a choice of conformal factor that makes the boundary totally geodesic in the enveloping manifold (there is much gauge freedom in choice of that conformal factor).

Similar examination is made of other boundaries, such as timelike infinity and timelike and spacelike singularities.  These are much simpler, as they admit a unique connection from a similar limiting process (i.e., no gauge freedom); and that connection is complete.
\endabstract

\endtopmatter

\document

\head 0. Introduction
\endhead 

It is common to consider a conformal boundary at infinity for many spacetimes, particularly static, spherically symmetric ones.  Typically, this boundary is topologically $\R1 \times \Sph2$, topological product of the line and the 2-sphere.  But of some question is whether that $\R1$-factor is ``naturally" really parametrized as a complete line, or as a finite or semi-infinite interval.  

The difference between viewing a topological line as, on the one hand, a complete line, or, on the other hand, as finite in one or both directions, is essentially a matter of geometry---if not metric geometry, then of an affine (or linear) connection, a way to measure acceleration of vector fields, even irrespective of a metric (Riemannian or otherwise).  The connection is complete if every geodesic has its affine parameter defined on $(-\infty, \infty)$---and this applies to $\R1 \times \Sph2$ as well as just  a line.

There are multiple challenges in looking for an appropriate affine connection on the boundary of a spacetime.  First, which boundary?  While it is common to embed a suitably symmetric spacetime $M$ into an ambient non-physical spacetime $M'$ in a conformal manner, using the boundary of $M$ in $M'$ as the boundary at infinity, there is no one natural way to do this---if nothing else, the geometry on the boundary is not uniquely determined; so one must be concerned about obtaining an answer to completeness on a non-universal boundary construction.  Also, there are spherically symmetric static spacetimes which simply do not admit a conformal boundary at all; these pose a problem for any process depending upon a conformal embedding.  This paper resolves this issue by looking at the future causal boundary, which has universal properties in terms of causal relations.  Of course, in classical spacetimes, this is topologically and causally the same as the conformal boundary which is generally used;  the point here is to rely only on universally defined properties of the boundary, and to use a boundary construction which exists for any strongly causal spacetime.

Once one is looking at a given boundary---and assuming one can put a natural differentiable structure on it---the question then becomes, what affine connection?  Boundary plus spactime form, one hopes, a manifold with boundary, and the spacetime bears the Levi-Civita connection from the spacetime metric; but this connection does not extend to the boundary, nor does the boundary at infinity bear any naturally occurring non-degenerate metric to yield its own Levi-Civita connection, if, as expected, it is future null infinity, i.e., naturally interpreted as a null submanifold.  So the matter of finding a naturally occurring affine connection is rather unclear.

This paper seeks a solution to that problem by looking at a limit of affine connections in ``nearby" null submanifolds.  While the choice of null submanifolds going out to infinity seems natural, there is no one natural choice of affine connection on each submanifold; rather, there is a family of choices, and different choices result in different limit connections on the boundary---or result in no limiting connection at all.  Thus we have a constellation of reasonable choices for a limit connection; particularly simple choices are complete, but others are not.

Another possible approach harks back to the conformal boundary:  One may look into the possibility of using a conformal embedding of the spacetime into a Lorentz manifold and examine whether there is an induced connection on the boundary of the image of the spacetime, due to that boundary being geodesically complete in the nonphysical spacetime; this does not exist for all spacetimes one may wish to consider, but it does for a great many.  The interesting aspect here is that the exact same constellation of connections is derived for $\Cal I^+$---future null infinity---interpreted as one of many choices of conformal boundary, as is derived in the previous manner.  So whether one prefers a (putatively) natural connection on $\Cal I^+$, understood as the future causal boundary, derived from (somewhat arbitrary) choices made on null submanifolds approaching $\Cal I^+$; or a connection derived from a (quite arbitrary) conformal embedding in a manner that is geodesically complete on $\Cal I^+$---assuming such exists---one arrives at the same end result, the same constellation of admissible connections on $\Cal I^+$.

Section 1 looks at the differential topology of the future causal boundary; this provides the groundwork for the main part of the paper.  The new material here is not the topology of the boundary itself, but explicating how it connects to the spacetime, enabling an extension of the differentiable structure of the spacetime to spacetime-{\it cum}\/-boundary.  Section 2 provides an analysis of the main idea for this paper: examining completeness of putatively natural connections on future null infinity of static, spherically symmetric spacetimes; both methods alluded to above are examined (i.e., connections on null hypersurfaces approaching the future causal boundary; and conformal boundaries that are geodesically complete in the non-physical spacetime). Section 3 looks at several other boundaries, such as timelike infinity in Anti-de Sitter, the timelike singularity in Reissner-Nordstr\"om, and the spacelike singularity in Schwarzschild, using similar means; the answer is much simpler: a unique connection on the boundary results, and it is complete.

\head 1. Future Null Infinity for Static, Spherically Symmetric Spacetimes: Differential Topology
\endhead

\subhead Causal boundary
\endsubhead

This section explores in detail the topology of the future causal boundary for static, spherically symmetric spacetimes---in particular, how the boundary connects to the spacetime---in such a way as to determine a natural extension to the boundary of the differentiable structure of the spacetime.

The causal boundary was introduced in 1973 in \cite{GKP}.  In 1998, \cite{H1} established a form of categorical universality for the future causal boundary constructions, in terms of creating a ``future-complete chronological set", i.e., just looking at a spacetime with boundary as a set with a chronology relation.  In 2000 \cite{H2} introduced a topology (the ``future-chronological topology") for any chronological set and showed how this produces a good topology for the future causal boundary in many contexts (though universality for the topology was established only for instances in which the causal boundary is spacelike); this is a different topology than the 1973 one, which was badly flawed (not yielding the intuitive topology even for the boundary of Minkowksi space). While there is still some uncertainty as to the best topology to use in general (see \cite{FHeSn1} for a discussion), there is good agreement on using the chronological topology for spacetimes similar to classical models such as standard static spacetimes, as exemplified in \cite{FHr} from 2007.  

In particular, the future-chronological topology applied to the future causal boundary for Minkowski space duplicates the topology from the standard conformal embedding of Minkowski space into the Einstein static
 spacetime (as seen, for instance, in \cite{HwE}). 

So why not just use the conformal boundary?  One reason is that there are static, spherically-symmetric spacetimes which do not admit a conformal boundary; but there is always the future causal boundary. 

\subhead Remarks on causal boundary and chronological topology generally
\endsubhead

The causal boundary, originally introduced in \cite{GKP}, has been refined and studied in more recent papers such as \cite{Hr1}, \cite{Hr2}, \cite{FHr}, \cite{FHeSn1}, and \cite{FHeSn1}.  The future causal boundary treats both spactime points and boundary points as IPs: indecomposable past sets, the pasts of causal curves; a boundary IP, or TIP, is the past of a future-endless causal curve.   Concentrating solely on the future causal boundary \cite{Hr1} showed that the process of adding that boundary to a strongly causal spacetime is a categorically universal process for producing a chronological set  that future-completes the spacetime.  Relevant nomenclature here: A {\it chronological set}\/ is a set with a chronology relation $\ll$; a {\it future chain}\/ in a chronological set is a sequence $p_1 \ll \cdots \ll p_n \ll p_{n+1} \ll \dots$; a {\it future limit}\/ $p$ for a future chain $\{p_n\}$ satisfies $I^-(p) = I^-[\{p_n\}]$ (where $I^-$ denotes chronological past); a chronological set is {\it future-complete}\/ if every future chain has a future limit; and for $M$ a strongly causal spacetime, $\hat M$ is $M$ together with the $\fcb(M)$, the future boundary points (TIPs), with a standard extension of the chronology relation $\ll$ on $M$ to $\hat M$.  {\it Universality}\/ here means (1) $i: M \to \hat M$ preserves $\ll$ and preserves future limits of future chains, (2) $\hat M$ is future-complete, and (3), for every map $f: M \to X$ into a future-complete chronological set with $f$ preserving $\ll$ and future limits of future chains, there is a unique extension $\hat f: \hat M \to X$ of $f$ (i.e., $\hat f \circ i = f$) so that $\hat f$ also preserves $\ll$ and future limits of future chains.  (In a general chronological set---i.e., not necessarily a spacetime---an IP is the past of a future chain; a boundary IP is the past of a future chain with no future limit.)

A topology---the future-chronological topology---was introduced in \cite{Hr2}, 
intending to give a reasonable topology for a great many chronological sets, in particular for $\hat M$ when $M$ is a modestly well-behaved spacetime; this topology misbehaves in some specially manufactured cases (the reason for further explorations of topology in \cite{FHeSn1}), but has excellent properties for a wide range of classical spacetime models, as shown in \cite{FHr}.  For a chronological set $X$, the {\it future-chronological topology}\/ is defined in terms of limits of sequences (a countability condition being employed to guarantee this is sufficient to define a topology).  For the common case in which the past of any point in $X$ is an IP, the limits are given thus:  For any sequence $\sigma = \{x_i\}$, the {\it future-chronological limit set}\/ $\hat L(\sigma)$ is defined to be those points $x$ such that (1) every element of $I^-(x)$ is in the past of all but finitely many $x_i$ and (2) $I^-(x)$ is a maximal IP among those IPs $Q$ satisfying, every point of $Q$ is in the past of infinitely many $x_i$.  Then a set $A$ in $X$ is closed in this topology iff for any sequence $\sigma$ in $A$, $\hat L(\sigma) \subset A$.  Among other desirable properties:  This recreates the manifold topology for $X$ when $X$ is any strongly causal spacetime; for any chronological set, the inclusion $i: X \to \hat X$ is continuous with open dense image; and with $X = \Min4$, we get the same result for $\hat X$ as adding the conformal boundary from the standard conformal embedding in the Einstein static spacetime (see, for instance, \cite{HE})---something famously not true for the original GKP topology.  This topology is always ${\text T}_1$ (points are closed), but need not be ${\text T}_2$ (Hausdorff); but this is a feature, not a bug, as non-Hausdorffness reveals relevant physical properties (see \cite{FHr}).  

In general, let $\hat X$ denote $X$ together with the future causal boundary, $\fcb(X)$; this has a chronology relation, extending that in $X$, defined thus, for $x$ in $X$ and $P$ and $Q$ IPs in $X$:  $x \ll P$ if $x \in P$; $Q \ll x$ if $Q \subset I^-(y)$ for some $y \ll x$; and $Q \ll P$ if $Q \subset I^-(y)$ for some $y \in P$.  We can also define a causality relation to add to any chronological set:  For $x$ and $y$ (for instance, in $\hat X$), define $x \prec y$ if $I^-(x) \subsetneq I^-(y)$.  There are spacetimes for which this will add additional causal relations, but they won't appear in spacetimes considered in this paper.

Here is a way to relate the future causal boundary (with the future-chronological topology) to reasonable conformal boundaries.  Let $M$ be a strongly causal spacetime; we will look at conformal mappings $\phi: M \to N$ where $N$ is a globally hyperbolic (non-physical) spacetime.  (The reason for restricting to globally hyperbolic targets is so that when, in $N$, $I^-(x) \subset I^-(y)$, we can conclude $x \prec y$, reflecting the definition of $\prec$ in the chronological set $\hat M$.  Briefly: For $I^-(x) \subset I^-(y)$ and $x \neq y$ in globally hyperbolic $N$, take any $x' \ll x$; then $x' \ll y$, and the relative compactness of $I^+(x') \cap I^-(y)$ shows the existence of a causal curve from $x'$ to $y$, through $x$.)  If $\phi$ is future-completing, i.e., provides a limit to the image of any future-directed timelike curve in $M$,  then Proposition 6 \cite{Hr1} yields a unique extension $\hat\phi: \hat M \to N$ that preserves $\ll$ and future limits of future chains; for $P$ an IP in $M$ this is defined by $\hat\phi(P) = \lim \phi(\gamma(t))$ where $\gamma$ is any causal curve whose past is $P$.  Let $\bar M_\phi$ denote $M$ together with the boundary $\partial_\phi(M)$, topologized by means of $\phi$ (i.e., $\bar M_\phi$ is identified with closure of $\phi(M)$ in $N$).  But while $\hat\phi: \hat M \to \bar M_{\phi}$ is ``future-continuous" (i.e., makes good sense in consideration of future chains), it is not necessarily continuous.

An appropriate question is, why restrict to just the future causal boundary---why not the past causal boundary and, particularly, an amalgamation of the two?  The Szabados relation, as seen, for instance, in \cite{FHeSn1}, provides a good basis for combining future and past causal boundary in many cases, including some that will be considered in this note; but there are considerable theoretical uncertainties in how to handle this amalgamation generally, most particularly with respect to appropriate topology.  (As an example of the theoretical difficulties:  While there is a categorical naturalness and uniqueness for the future causal boundary---more specifically, for the process of future-completion---in respect of future-continuous maps, there does not appear to be any similar categorical naturalness or uniqueness for the amalgamation of future and past causal boundary; see \cite{Hr1} for a categorical treatment of future-completion and an example of failure to find a boundary which simultaneously future- and past-completes for a particular spacetime.)  For the spacetimes considered in this note, future and past causal boundaries are mirrors of one another, and there is nothing particularly to be gained by adding consideration of the past causal boundary or the obvious (and Szabados-inspired) amalgamation of the two.  For consideration of more general spacetimes, it would be appropriate to look more deeply into the amalgamation, whether by the methods in \cite{FHeSn1} or other possibilities.

The balance of this note will, therefore, be concerned with the future causal boundary using the future-chronological topology.

\subhead Causal boundary for static, spherically-symmetric spacetimes
\endsubhead

What is the general form for a static, spherically-symmetric spacetime?  By results in \cite{Hr3} (discussion above Proposition 2.11), any static spacetime $M$ with complete timelike Killing field can be represented as a group quotient of a product spacetime, $(\Bbb R^1 \times \tilde Q)/G$, where $\tilde Q$ is the universal cover of the space $Q$ of static observers for $M$, $\Bbb R^1 \times \tilde Q$ is a standard static spacetime (i.e., has metric $ds^2 = -(\Omega \circ \pi)dt^2 + \pi^*\tilde h$ for $\pi$ being projection to $\tilde Q$, $\tilde h$ a Riemannian metric on $\tilde Q$, and $\Omega: \tilde Q \to \Bbb R^+$ the length-squared of the Killing field), $G = \pi_1(M) = \pi_1(Q)$ (the fundamental group), with $G$ acting on the product via $a \cdot(t,q) = (t + \rho(a), a \cdot q)$ for some group representation $\rho: G \to  \Bbb R$; causal structure in $M$ is determined by the nature of the group representation, as explained in \cite{Hr3}.  For a spherical symmetry we generally mean that the static rest-space (i.e., the space of static observers) should have a radially dependent geometry on the product $Q = (r_1,r_2) \times\Sph2$, where $(r_1,r_2)$ is any interval in the real line; thus, the metric on $Q$ is $h = c(r)^2dr^2 + b(r)^2 h_{\Sph2}$ for some functions $c, b: (r_1,r_2) \to \Bbb R^+$, with $h_{\Sph2}$ denoting the standard metric on the unit 2-sphere (most typically, $b(r) = r$).  Since $Q$ is simply connected, we have $G$ is necessarily the trivial group, and $M$ is itself standard static.  We will also want $\Omega$ to be spherically symmetric, i.e., be independent of the $\Sph2$ factor in $Q$.  Thus we will be considering representation
$$\align
M & = \Bbb R^1 \times (r_1,r_2) \times \Sph2 \\
ds^2 = g & = -\Omega(r)dt^2 + c(r)^2dr^2 + b(r)^2h_{\Sph2}
\endalign$$
But as the causal boundary is conformally invariant, we can, for purposes of topology and causality, alternatively
 consider a conformally related metric $\bar g = \frac1\Omega g$, resulting in the representation
$$\align
M & = \Bbb R^1 \times (\alpha,\omega) \times \Sph2 \\
\bar g & = -dt^2 + d\rho^2 + a(\rho)^2h_{\Sph2}
\endalign$$
where we pick some $r_0 \in (r_1,r_2)$, define $\bar c = \frac1{\sqrt\Omega}c$, and then $\alpha = \int_{r_1}^{r_0} \bar c$, $\omega = \int_{r_0}^{r_2} \bar c$, $\rho(r) = \int_{r_0}^r \bar c$, and $a(\rho(r)) =\frac1{\sqrt{\Omega(r)}}b(r)$.

If the function $a(\rho)$ has simple behavior, then by Theorems 6.2 and 6.7 and discussion in Example 6.1.3 of \cite{FHr}, we know exactly what the future completion of $M$ is, using the future-chronological topology:  $\hat\partial(M)$ is a cone over $\partial_B(Q)$ (the Busmen boundary of $Q$, so called as it is formed from Busmen functions of all endless unit-speed curves); and $\partial_B(Q)$ consists of two disjoint spaces, $B_\alpha$ and $B_\omega$, each either a point or a 2-sphere: for $\iota$ being either $\alpha$ or $\omega$, $B_\iota$ is a point if $\int_{\rho_0}^\iota 1/a(\rho)^2 d\rho$ is infinite (for some choice of $\rho_0 \in (\alpha,\omega)$), while $B_\iota$ is a sphere if that integral is finite.  We need to know how  $\hat\partial(M)$ attaches to $M$ to form $\hat M$.  This will take the form of the theorem immediately below; in brief, the attachment is essentially that of the attachment of future null infinity in $\Min2$.  (Theorems 6.2 and 6.7 of \cite{FHr} explicate only the topology of $\fcb(M)$, isolated from $\hat M$; however, the proof of Proposition 6.1 is applicable to all of $\hat M$, and that is what we will be using to understand how $\fcb(M)$ attaches to $M$ in $\hat M$.)

We also need to know the causal structure in $\hat M$:  The boundary cone elements, i.e., the lines that are associated to each point in $B_\iota$, are all timelike if  $\iota$ is finite, and all null if $\iota$ is infinite.  Causal relations among the boundary points and $M$ are just what are intuitively obvious from the product structure on $M$, and there are no further causal relations among the boundary points (contrary to what can happen in more complex situations as in Example 2.1 of \cite{FHr}).

These same results on $\hat M$ apply for any compact Riemannian manifold $K$ in place of $\Sph2$.  In case $K$ is not simply connected, a static spacetime with static rest space a warped product $Q = (r_1,r_2) \times_a K$ might not be standard static, but a quotient of $\Bbb R^1 \times \tilde Q$; that can complicate the future completion $\hat M$ (see \cite{Hr4}).  But we won't consider that complication in this paper, looking only at standard static spacetimes.

The required simple behavior of the function $a(\rho)$ is that as $\rho$ approaches either end of the interval, $a(\rho)$ eventually be monotonic increasing towards that end, that is to say, decreasing on the $\alpha$-end and increasing on the $\omega$-end (this is typically the case for the boundary at physical infinity, as well as some cases of event horizon); an alternative behavior that also works, at least for the case of the integral condition being infinite, is that $a(\rho)$ decrease monotonically to 0 near an end of the interval (this is typically the case for a physical ``center" without horizon or singularity). 

To look at the exact topological structure of $\hat M$ we will first need to look at the simple two-dimensional spacetime $\Minn = \Min1 \times (\alpha,\omega)$.  We can represent this with standard time and space co\"ordinates $(t,\rho) \in \R1\times(\alpha,\omega)$; or with straight-forward null co\"ordinates $\lambda = t - \rho$, $\mu = t + \rho$ and $(\lambda,\mu)$ satisfying $\mu - 2\omega < \lambda < \mu - 2\alpha$ for any $\mu \in \R1$; or with compactified null co\"ordinates $\eta = \tan^{-1}\lambda$, $\sigma = \tan^{-1}\mu$ and $(\eta,\mu)$ satisfying $\tan^{-1}(\tan\sigma - 2\omega) < \eta < \tan^{-1}(\tan\sigma - 2\alpha)$ for any $\sigma \in (-{\frac\pi2},{\frac\pi2})$.  Proposition 6.1 of \cite{FHr} (with Riemannian factor $(\alpha,\omega)$) shows that we can form the future completion and the future causal boundary of $\Minn$ very simply; it is easily observed using the obvious embedding of $\Minn$ into $\R2$ by using the compactifying co\"ordinates, $(\eta,\sigma) \mapsto (\eta,\sigma)$, that is to say, $(\R1 \times (\alpha,\omega), -(dt)^2 + (d\rho)^2) \to ((-{\frac\pi2},{\frac\pi2}) \times (-{\frac\pi2},{\frac\pi2}),-(\sec^2\eta\sec^2\sigma) \,d\eta d\sigma)$ via $(t,\rho) \mapsto (\eta(t,\rho), \sigma(t,\rho)$.  (Proposition 6.1 of \cite{FHr} allows us to conclude that the future chronological topology on $\widehat{\Minn}$ is given by representing each point $p = (t, \rho)$ of $\Minn$ as the function on $(\alpha,\omega)$ given by $x \mapsto t - |x - \rho|$, i.e., the function whose graph in $\R1\times (\alpha,\omega)$ is the boundary of $I^-(p)$; representing each TIP $P$ by the linear function on $(\alpha,\omega)$ whose graph is the boundary of $P$, except for the TIP $i^+$ which is represented by the function of constant value $\infty$; and using the function-space topology on those functions as the topology on $\widehat{\Minn}$. That function-space topology is easily seen to be preserved by the co\"ordinates of points in the embedding mentioned.)  In the usual case of $\Min2$---i.e., $(\alpha,\omega) = (-\infty,\infty)$---we have the boundary of the image of $\Min2$ consisting of a single point of timelike infinity $i^+ = (\frac\pi2,\frac\pi2)$; two components ($\sigma = \frac\pi2$ and $\eta = \frac\pi2$) of future null infinity, $\Cal I^+ = [(-\frac\pi2,\frac\pi2)\times \{\frac\pi2\}] \cup [\{\frac\pi2\}\times(-\frac\pi2,\frac\pi2)]$; two points of spatial infinity $i^0 = \{(-\frac\pi2,\frac\pi2),(\frac\pi2,-\frac\pi2)\}$ (frequently commingled into a single point); and, similarly, two components of past null infinity $\Cal I^-$ and one point of past timelike infinity $i^-$.  In case $\alpha$ or $\omega$ is finite, the boundary on the respective side is a single timelike curve between the points we might as well still call $i^-$ and $i^+$ (though usages such as ``timelike infinity" and the like lose explanatory power).

It will be helpful to be explicit about various portions of the future causal boundary for $(M,\bar g) = (\R1 \times (\alpha,\omega) \times K, -dt^2 + d\rho^2 + a(\rho) h_K)$ with $K$ compact and $a(\rho)$ decreasing on the $\alpha$-end of the interval (i.e., increasing with movement towards $\alpha$) and increasing on the $\omega$-end.  As already mentioned, we know from \cite{FHr} that $\fcb(M)$ is topologically a cone on the disjoint union $B_\alpha \cup B_\omega$, where $B_\iota$ is either a point or $K$; and the cone-elements on $B_\iota$ are either timelike or null, according as $\iota$ is finite or infinite, with the vertex of the cone being the future endpoint.  In fact, that vertex is the TIP which is all of $M$ (thus, it is to the future of every point of $M$); let us denote it $i^+$.  As our intent is to put a differentiable structure on the boundary, we'll have to omit this point from consideration in order to obtain a manifold; so let $\fcb^0(M) = \fcb(M) - \{i^+\}$, which is the disjoint union of two lines, or of a line and $\R1\times K$, or of two copies of $\R1\times K$: plainly a candidate for a differentiable manifold. Let us call these two components---each a cylinder on $B_\alpha$ or $B_\omega$---respectively $\fcb^0_\alpha(M)$ and $\fcb^0_\omega(M)$.  In the degenerate case of $\Minn$ (technically, $K$ is empty) the nomenclature still works: For $\omega$ finite, $\fcb^0_\omega(\Minn)$ is a timelike curve, future-directed from $i^-$ to $i^+$; for $\omega = \infty$, $\fcb^0_\infty(\Min2_{(\alpha,\infty)})$ is a null line, future-directed from (one of the possibly two points of) $i^0$  to $i^+$.

\proclaim{Theorem 1.1} Let $M$ be a standard static spacetime of the form 
$$\align
M & = \R1 \times (\alpha,\omega) \times K \\
ds^2 = \bar g & = -dt^2 + d\rho^2 + a(\rho)^2h_K
\endalign$$
where $K$ is a compact manifold with Riemannian metric $h_K$ ($\alpha$ and $\omega$ are allowed to be infinite).  

Suppose $a(\rho)$ is decreasing on some interval $(\alpha,\rho_-]$ and increasing on some interval $[\rho_+,\omega)$.  For $\iota$ being either $\alpha$ or $\omega$, let $B_\iota = K$ if $\int_{\rho_+}^\iota \frac1{a(\rho)^2} \,d\rho$ is finite, and otherwise $B_\iota$ is a point, *.  

Alternatively, suppose $a(\rho)$ is increasing on some interval $(\alpha,\rho_-]$ with $a(\rho) \to 0$ as $\rho \to \alpha$, and increasing on some interval $[\rho_+,\omega)$.  Let $B_\omega$ be as in the paragraph above, but only allow $\int_\alpha^{\rho_-} \frac1{a(\rho)^2} \,d\rho = \infty$ and $B_\alpha = *$.  (Or reverse the roles, with $a(\rho)$ decreasing on $(\alpha,\rho_-]$ and decreasing on $[\rho_+,\omega)$ with $a(\rho) \to 0$ as $\rho \to \omega$; and with $B_\alpha$ as in the paragraph above, but with $\int_{\rho_+}^\omega \frac1{a(\rho)^2} \,d\rho = \infty$ and $B_\omega = *$.)

Then in the future-chronological topology,
\roster
\item $\fcb(M$) has the topology of a cone on the disjoint union $B_\alpha \cup B_\omega$  \newline with cone-elements on $B_\iota$ timelike if $\iota$ is finite, otherwise null, \newline with the cone vertex at the future end, and
\item $\hat M$ has the topology of $(\widehat{\Minn}\times K)/\sim$ 
\newline where $\sim$ is the equivalence relation which smashes the $K$-factor to a point at $\{i^+(\Minn)\}\times K$ and, if either $B_\iota$ is a point, also along $\fcb^0_\iota(\Minn) \times K$.
\endroster
In other words:  $\fcb(M)$ is attached to $M$ in $\hat M$ as $\fcb(\Minn)$ is attached to $\Minn$ in $\widehat{\Minn}$.
\endproclaim

\demo{Proof}
Conclusion (1) is precisely Theorem 6.7 of \cite{FHr} (or, in the alternative, the discussion in section 6.1.3 devoted to the singularity in Reissner-Nordtr\"om)) and is included here
 just for completeness.  Our burden here is to establish conclusion (2) as well.

Proposition 6.1 of \cite{FHr} gives conditions for any standard static spacetime $M = \Min1 \times Q$, under which the future-chronological topology can be understood as just the function-space topology applied to functions on $Q$ whose graphs are the boundaries of IPs in $M$---specifically, Busemann functions associated to unit-speed curves in $Q$---and that the only unit-speed curves that need to be considered are geodesics which are the limits of minimizing arcs going out to infinity.  Actually, the statement of the proposition mentions only the future causal boundary of the spacetime, but the proof plainly applies to the entire future completion $\hat M$.

Theorem 6.2 of \cite{FHr} establishes that standard static spactimes of the form considered here obey the hypotheses of Proposition 6.1.  Thus, all we need do is examine the Busemann functions for limit geodesics in $Q = (\alpha,\omega) \times K$---which amounts to saying, radial geodesics, constant in $K$---and show that the function-space topology on those functions yields the topology claimed in (2).

We need to consider only one component of $\fcb^0(M)$ at a time---say, $\fcb^0_\omega(M)$.  So the curves whose Busemann functions we need consider are curves terminating at points of $Q$ (generating IPs which are not boundary points) and curves in $Q$ of the form $s \mapsto (s,x)$ for $x \in K$ (generating boundary IPs).  Now, for any unit-speed curve $c: [0,T) \to Q$, the associated Busemann function is $b_c: Q \to \R{}$ defined by $b_c(q) = \lim_{s \to T}(s - d(c(s),q))$ (where choice of starting parameter for $c$ amounts to adding a constant to the function).  Thus, for Busemann functions representing points $(\rho_0,x_0)$ we have functions of the form $\delta^{t_0}_{\rho_0,x_0} : (\rho,x) \mapsto t_0 - d((\rho,x),(\rho_0,x_0)))$; and for Busemann functions generated from radial curves going out to $\rho = \omega$, we have functions of the form $b^{t_0}_{x_0} : (\rho,x) \mapsto t_0 + \lim_{s \to \omega}(s - d((\rho,x),(s,x_0)))$.  The corresponding functions for $\Minn$ (defined on $(\alpha,\omega)$) are $\delta^{t_0}_{\rho_0} : \rho \mapsto t_0 -|\rho - \rho_0|$ and $b^{t_0}: \rho \mapsto t_0 + \lim_{s \to \omega}(s - |s - \rho|) = t_0 + \rho$.  In both spaces we also have $i^+$ represented by the function which is constantly $\infty$ (this is the only Busemann-type function which has any infinite values; in particular, the function constantly $-\infty$ is not considered, as that would represent the empty set as a supposed IP).  We need to show that sequences $\{\delta^{t_n}_{\rho_n,x_n}\}$ have the same form of point-wise convergence to boundary functions $\{b^{t_0}_{x_0}\}$ and $i^+$ as do the sequences $\{\delta^{t_n}_{\rho_n}\}$ to boundary functions $b^{t_0}$ and $i^+$, also taking into account convergence in $K$, in case $B_\omega = K$.  This comes down to showing the following (where $I = \int_{\rho_0}^\omega \frac1{a(\rho)^2} d\rho$ for some $\rho_0 \in (\alpha,\omega)$), in terms of approach to $i^+$, of approach to the boundary at $\rho = \omega$, and of failure to converge:
\roster
\item if $\{t_n - |\rho_n|\} \to \infty$, then $\{\delta^{t_n}_{\rho_n,x_n}\} \to i^+$ (i.e., $\infty$)
\item if $\{t_n - \rho_n\} \to t_\infty$ (finite), then 
\newline (a) if $I = \infty$ and $\{\rho_n\} \to \omega$, then $\{\delta^{t_n}_{\rho_n,x_n}\} \to b^{t_\infty}_*$ (for $*$ any point in $K$)
\newline (b) if $I < \infty$ and $\{\rho_n\} \to \omega$ and $\{x_n\} \to x_\infty$, then $\{\delta^{t_n}_{\rho_n,x_n}\} \to b^{t_\infty}_{x_\infty}$
\newline (c) if $I < \infty$ and $\{\rho_n\} \to \omega$ and $\{x_n\}$ has no limit, then $\{\delta^{t_n}_{\rho_n,x_n}\}$ has no limit in $\fcb(M)$
\newline (d) if all $\rho_n \le \rho_\infty < \omega$, then $\{\delta^{t_n}_{\rho_n,x_n}\}$ has no limit in $\fcb^0_\omega(M) \cup \{i^+\}$
\item if $\{t_n - |\rho_n|\} \to -\infty$, then $\{\delta^{t_n}_{\rho_n,x_n}\}$ has no limit in $\fcb(M)$
\endroster
(The case of $\{t_n + \rho_n\} \to t_\infty$ is that of $\{\delta^{t_n}_{\rho_n}\}$ approaching an element of $\fcb^0_\alpha(\Minn)$ rather than $\fcb^0_\omega(\Minn)$, so would fall under the examination of $\fcb^0_\alpha(M)$---entirely analogous.)

We repeatedly will use the following elementary facts about distance $d$ in $Q$,  (with $d_K$ denoting distance in $K$ and $a(\rho)$ being increasing on $[\rho_+,\omega)$):
$$\align
d((\rho,x),(\rho',x')) & \ge |\rho -\rho'| \\
d((\rho,x),(\rho',x')) & \ge a(\rho)d_K(x,x') \\
& \quad (\text{assuming }\rho_+ \le \rho \le \rho') \\
d((\rho,x),(\rho',x')) & \le \rho' - \rho + a(\rho)d_K(x,x') \\
& \quad (\text{assuming }\rho_+ \le \rho \le \rho')
\endalign$$ 
(In the alternate hypotheses, with, say, $a$ decreasing on $[\rho_+,\omega)$, we must replace $a(\rho)$ with $a(\rho')$ in the second line; but the third line is true with either $a(\rho)$ or $a(\rho')$.)

(1): We assume $\{t_n - |\rho_n|\} \to \infty$; we want to show $\{\delta^{t_n}_{\rho_n,x_n}\} \to \infty$.

For all $(\rho,x) \in (\alpha,\omega) \times K$, 
$$\align
\dn(\rho,x) & = t_n - d((\rho,x),(\rho_n,x_n))\\
& \ge t_n - |\rho - \rho_n| \\
& \ge t_n - |\rho_n| - |\rho| \\
& \to \infty 
\endalign$$

(2): Throughout this section, we can write $t_n = \rho_n + t_\infty + \epsilon_n$, where $\{\epsilon_n\} \to 0$.

(2a,b): We further assume $\{\rho_n\} \to \omega$ and, in case $I < \infty$, $\{x_n\} \to x_\infty$; we want to show $\{\delta^{t_n}_{\rho_n,x_n}\} \to b^{t_\infty}_{x_\infty}$ (in case $I = \infty$, any $x_\infty$ will serve).

Define the functions $\delta^- = \liminf_{n \to \infty}\dln$ , $\delta^+ = \limsup_{n \to \infty}\dln$. First note $\delta^- \neq -\infty$, i.e., for any $(\rho,x) \in (\alpha,\omega) \times K$, $\{t_n - d((\rho_n,x_n),(\rho,x))\}$ has a finite infimum as $n \to \infty$:  We see this is true by observing that for $n$ sufficiently large (so $\rho_n > \rho$), 
$$\align
d((\rho_n,x_n),(\rho,x)) & \le \rho_n - \rho + a(\rho)d(x_n,x) \\
& \le \rho_n - \rho + a(\rho)\text{diam}(K)
\endalign$$
so
$$\align
t_n - d((\rho_n,x_n),(\rho,x)) & \ge t_n - \rho_n + \rho - a(\rho)\text{diam}(K) \\
& = t_\infty + \epsilon_n - a(\rho)\text{diam}(K)
\endalign$$
and $\delta^-(\rho,x) \ge t_\infty - a(\rho)\text{diam}(K) > -\infty$.  (This is the only passage in the proof that is potentially sensitive to the alternate hypotheses in the theorem.  But also with the alternative hypothesis of $a$ decreasing on $[\rho_+,\omega)$, we still have $d((\rho_n,x_n),(\rho,x)) \le \rho_n - \rho + a(\rho)d(x_n,x)$, since $a(\rho) \ge a(\rho_n)$ for $n$ sufficiently large; thus, everything goes through.)

Next note $\delta^+ \neq \infty$:  For $n$ sufficiently large, $d((\rho_n,x_n),(\rho,x)) \ge \rho_n - \rho$, so
$$\align
t_n - d((\rho_n,x_n),(\rho,x)) & \le t_n - \rho_n + \rho \\
& = t_\infty + \epsilon_n + \rho
\endalign$$
and $\delta^+(\rho,x) \le t_\infty + \rho < \infty$.  By Corollary 5.13 of \cite{FHr}, knowing $\delta^-$ is finite tells us that  some subsequence of $\{\dln\}$ approaches some point $b \in \fcb(M)$ (a unique point, as, by Theorem 6.2 of \cite{FHr}, $\fcb(M)$ is hausdorff); since $\delta^+$ is finite, we know that $b$ is not $i^+$, so it is some $b^{\bar t}_{\bar x} \in \fcb^0_\omega(M)$; and, furthermore (again from Corollary 5.13), for all $t < \delta^-(\rho,x)$, $b^{\bar t}_{\bar x} \ge \delta^t_{\rho,x}$.  We will then be finished if we show $\bar t = t_\infty$ and, in case $I = \infty$, $\bar x = x_\infty$, since every subsequence will then have a convergent subsubsequence with the same limit.

To show $\bar t = t_\infty$:  We have, for all $(\rho, x)$,
$$\align
\lim_{k\to\infty}\delta^{t_{n_k}}_{\phi_{n_k},x_{n_k}}(\rho,x) & = b^{\bar t}_{\bar x}(\rho,x) \\
\lim_{k\to\infty}\(t_{n_k} - d((\rho_{n_k},x_{n_k}),(\rho,x))\) & = \bar t +  \lim_{s\to\infty}\(s - d((s,\bar x),(\rho,x))\) \\
\lim_{k\to\infty}\(t_\infty + \rho_{n_k} + \epsilon_{n_k} - d((\rho_{n_k},x_{n_k}),(\rho,x))\) & = \bar t + \lim_{k\to\infty}\(\rho_{n_k} - d(\rho_{n_k},\bar x),(\rho,x))\) \\
t_\infty + \lim_{k\to\infty}\( \rho_{n_k} - d((\rho_{n_k},x_{n_k}),(\rho,x))\) & = \bar t + \lim_{k\to\infty}\(\rho_{n_k} - d(\rho_{n_k},\bar x),(\rho,x))\) 
\endalign$$
hence
$$\bar t - t_\infty = \lim_{k\to\infty}\(d((\rho_{n_k},\bar x),(\rho,x)) - d((\rho_{n_k},x_{n_k}),(\rho,x))\) \tag{*}$$
Applying (*) to $x = \bar x$ we get (for $k$ sufficiently large)
$$\bar t - t_\infty = \lim_{k\to\infty}\(\rho_{n_k} - \rho - d((\rho_{n_k},x_{n_k}),(\rho,x))\) $$
and since $d((\rho_{n_k},x_{n_k}),(\rho,x)) \ge \rho_{n_k} - \rho$, we get $\bar t \le t_\infty$.  Then we consider $x = x_\infty$:  We know $d((\rho_{n_k},\bar x),(\rho,x_\infty)) \ge \rho_{n_k} - \rho$ and $d((\rho_{n_k},x_{n_k}),(\rho,x_\infty)) \le \rho_{n_k} - \rho + a(\rho)d(x_{n_k},x_\infty)$; then from (*) we derive
$$\align
\bar t - t_\infty & \ge \lim_{k\to\infty} -a(\rho)d(x_{n_k},x_\infty)\\
& = 0
\endalign$$
so $\bar t = t_\infty$.

To show $\bar x = x_\infty$ when $I < \infty$:  Suppose $\bar x \neq x_\infty$.  From Theorem 6.2 of \cite{FHr} we know $b^{t_\infty}_{\bar x} \neq b^{t_\infty}_{x_\infty}$: these two functions differ on some point $(\hat\rho,\hat x)$.  So we have 
$$\align
t_\infty + \lim_{s\to\omega}(s - d((s,\bar x),(\hat\rho,\hat x))) & \neq 
t_\infty + \lim_{s\to\omega}(s - d((s,x_\infty),(\hat\rho,\hat x))) \\
\lim_{s\to\omega}(d((s,x_\infty),(\hat\rho,\hat x)) & - d((s,\bar x),(\hat\rho,\hat x))) \neq 0
\endalign$$
or
$$\lim_{k\to\infty}(d((\rho_{n_k},x_\infty),(\hat\rho,\hat x)) - d((\rho_{n_k},\bar x),(\hat\rho,\hat x))) \neq 0 \tag{**}$$
Now, in part II.b of the proof of Theorem 6.2 of \cite{FHr}, it is shown that for any $\rho, \bar\rho \in (\alpha,\omega)$ and $x, y, z \in K$, $|d((\rho,x),(\bar\rho,y)) - d((\rho,x),(\bar\rho,z))| \le a(\rho)d(y,z)$.  This gives us
$$|d((\rho_{n_k},x_{n_k}),(\hat\rho,\hat
 x)) - d((\rho_{n_k},x_\infty),(\hat\rho,\hat x))| \le a(\hat\rho)d(x_{n_k},x_\infty)$$
so
$$\lim_{k\to\infty}(d((\rho_{n_k},x_{n_k}),(\hat\rho,\hat x)) - d((\rho_{n_k},x_\infty),(\hat\rho,\hat x))) = 0$$
Combining that with (**) yields
$$\lim_{k\to\infty}(d((\rho_{n_k},x_{n_k}),(\hat\rho,\hat x)) - d((\rho_{n_k},\bar x),(\hat\rho,\hat x))) \neq 0$$
and that violates (*) for $(\rho,x) = (\hat\rho,\hat x)$ when we recall $\bar t = t_\infty$.  Therefore, $\bar x = x_\infty$.

(2c).  Besides $\{t_n - \rho_n\} \to t_\infty$, we assume $I < \infty$, $\{\rho_n\} \to \omega$, and $\{x_n\}$ has no limit in $K$; we want to show $\{\dln\}$ has no limit in $\fcb(M)$.  Since $K$ is compact, there are subsequences $\{n_k\}$ and $\{m_k\}$ with $\{x_{n_k}\} \to x_\infty$ and $\{x_{m_k}\} \to y_\infty$ for two points $x_\infty \neq y_\infty$ in $K$.  From part (2a) or (2b) of this theorem, we know $\{\delta^{t_{n_k}}_{\rho_{n_k},x_{n_k}}\} \to b^{t_\infty}_{x_\infty}$ and $\{\delta^{t_{m_k}}_{\rho_{m_k},x_{m_k}}\} \to b^{t_\infty}_{y_\infty}$; and by Theorem 6.2 of \cite{FHr}, these are distinct points in $\fcb(M)$.  Therefore there is no limit of $\{\dln\}$.

(2d) Besides $\{t_n - \rho_n\} \to t_\infty$, we assume for some $\rho_\infty < \omega$, for all $n$, $\rho_n \le \rho_\infty$; we wish to show $\{\dln\}$ has no limit in $\fcb^0_\omega(M) \cup \{i^+\}$. We consider two cases.

i) $\omega = \infty$: First note that any Busemann function $b^{t_0}_{x_0}$ is unbounded above: $b^{t_0}_{x_0}(\rho,x_0) = t_0 + \lim_{s \to \infty}(s - d((\rho,x_0),(s,x_0))) = t_0 + \lim_{s \to \infty}(s- (s - \rho)) = t_0 + \rho$, which is unbounded as $\rho \to \infty$.  Now using $t_n = \rho_n + t_\infty + \epsilon_n$, we see that for any $\rho$ and $x$, $\dn(\rho,x) = t_n - d((\rho,x),(t_n,x_n)) = t_\infty + \rho_\infty + \epsilon_n - d((\rho,x),(t_n,x_n)) \le t_\infty + \rho_\infty + \epsilon_0$ for any choice of $\epsilon_0$ greater than all $\{\epsilon_n\}$.  Clearly this means $\dn(\rho,x)$ cannot approach $\infty$ (i.e. $\{\dn\}$ does not approach $i^+$); but it also means that $\{\dn\}$ cannot approach any $b^{t_0}_{x_0}$, as $\{\dn\}$ are uniformly bounded above by $t_\infty + \rho_\infty + \epsilon_0$, while $b^{t_0}_{x_0}$ is unbounded above. 

ii) $\omega < \infty$:  First suppose there is some $\rho_{-\infty} > \alpha$ such that for all $n$, $\rho_n \ge \rho_{-\infty}$.  Then for some subsequence $\{x_{n_k}\}$ has a limit $x_0$ and $\{\rho_{n_k}\}$ has a limit $\rho_0$.  Therefore, $\{\delta^{t_{n_k}}_{\rho_{n_k},x_{n_k}}\} \to \delta^{t_0}_{\rho_0,x_0}$ for $t_0 = \rho_0 + t_\infty$; but this is in $M$ and not in $\fcb(M)$, which is closed; therefore $\{\dln\}$ has no limit in $\fcb(M)$.  Now suppose there is some subsequence with $\{\rho_{n_k}\} \to \alpha$.  Then by analogy with (2a,b), we know $\{\delta^{t_{n_k}}_{\rho_{n_k},x_{n_k}}\}$ has a limit in $\fcb^0_\alpha(M)$; thus, $\{\dln\}$ has no limit in $\fcb^0_\omega(M) \cup \{i^+\}$.

(3) We assume $\{t_n - |\rho_n|\} \to -\infty$; we want to show $\{\dln\}$ has no limit in $\fcb(M)$.  For any $(\rho, x)$, we have $\dln(\rho,x) = t_n - d((\rho_n,x_n), (\rho, x)) \le t_n - |\rho_n - \rho| \le t_n - |\rho_n| + |\rho|$; therefore, $\{\dln(\rho,x)\} \to - \infty$, and $\{\dln\}$ has no limit in $\fcb(M)$.

\qed

\enddemo

The topology of the future causal boundary is neatly summarized for some of the most common model spacetimes in Corollary 6.8 of \cite{FHr}:

\proclaim{Common Model Spacetime Paradigm} Let $M$ be a spacetime with topology $\Bbb R^1 \times (r_{\text{min}},r_{\text{max}}) \times \Sph2$ (with $r_{\text{min}} > 0$) and metric $ds^2 = g = -f(r)dt^2 + \frac1{f(r)}dr^2 + r^2 h_{\Sph2}$ for $f(r) > 0$ on $(r_{\text{min}},r_{\text{max}})$.  If $\frac{r^2}{f(r)}$ is eventually increasing toward each end of the domain interval, $\hat M$ is a cone on $[r_{\text{min}},r_{\text{max}}] \times \Sph2$ (i.e., $(-\infty,\infty] \times [r_{\text{min}},r_{\text{max}}] \times \Sph2$, with $\{\infty\} \times [r_{\text{min}},r_{\text{max}}] \times \Sph2$ squashed to a point).  The boundary at $r = r_{\text{max}}$ is a null cone on $\Sph2$ if $\int_{r_0}^{r_{\text{max}}} \frac1{f(r)}\, dr = \infty$ (for $r_0$ any number in the domain interval), otherwise a timelike cone; and similarly for the boundary at $r = r_{\text{min}}$.

If $M = \Bbb R^1 \times (0,r_{\text{max}}) \times \Sph2$ with metric as above, and if $\frac{r^2}{f(r)}$ is eventually decreasing to 0 at the $r = 0$ end of the interval and eventually increasing at the other end, then the only change from above is the cone on $\Sph2$ at $r = 0$ is replaced by a cone on a point (i.e., a single line)---null if $\int_0^{r_0} \frac1{f(r)}\, dr = \infty$, otherwise timelike.

In all cases above, the topology is the indicated product topology, and the causal relations in $\hat M$ (in addition to what has been specified) are just the obvious ones from a product relationship.
\endproclaim

Thus the common notion of boundary at null (or even timelike) infinity for a spherically symmetric static spacetime is exemplified by a null (or timelike) cone on $\{r_{\text{max}}\}\times\Sph2$, as embedded in the spacetime treated as a timelike cylinder on $(r_{\text{min}},r_{\text{max}}) \times \Sph2$; typically $r_{\text{max}} = \infty$.  But timelike cones and timelike lines can also occur.

Some typical elements of this paradigm (details in \cite{FHr}):

\subhead Minkowski Space \endsubhead $(r_{\text{min}},r_{\text{max}}) = (0,\infty)$, $f(r) = 1$; boundary at $\infty$ is future null cone at infinity, boundary at $0$ is just the timelike line filling in the manifold.
 
\subhead External Schwarzschild \endsubhead 
$(r_{\text{min}},r_{\text{max}}) = (2m, \infty)$, $f(r) = 1 - \frac{2m}r$ (for $m > 0$); boundary at $\infty$ is future null cone at infinity, boundary at $2m$ is null cone at the event horizon for the singularity.

\subhead Medial Schwarzschild-de Sitter \endsubhead 
$(r_{\text{min}},r_{\text{max}}) = (r_-,r_+)$, $f(r) = 1 - \frac{2m}r - \frac13\Lambda r^2 $ (for $0 < \Lambda < \frac1{9m^2}$; $f(r) > 0$ on $(r_-,r_+)$); boundaries at $r_-$ and $r_+$ are null cones at event horizons---one at $r_-$ for a singularity and one at $r_+$ for spatial infinity.

\subhead External Schwarzschild--Anti-de Sitter \endsubhead
$(r_{\text{min}},r_{\text{max}}) = (r_0,\infty)$, $f(r) = 1 - \frac{2m}r - \frac13\Lambda r^2 $ (for $\Lambda < 0$; $f(r) > 0$ on $(r_0,\infty)$); boundary at $\infty$ is timelike cone at infinity, boundary at $r_0$ is null cone at event horizon for the singularity.

\subhead External and Internal Undercharged Reissner-Nordstr\"om \endsubhead 
$(r_{\text{min}},r_{\text{max}}) = (r_+,\infty)$ for External, $(0,r_-)$ for Internal, $f(r) = 1 - \frac{2m}r + \frac{q^2}{r^2}$ ( for $0 < |q| < m$; $f(r) > 0$ on $(0,r_-)$ and $(r_+,\infty)$); External: boundary at $\infty$ is  null cone at infinity, boundary at $r_+$ is  null cone at the outer event horizon; Internal: boundary at $r_-$ is  null cone at the inner event horizon, boundary at $0$ is a timelike line (the singularity). (Critically Charged and Overcharged Reissner-Nordstr\"om are also in this paradigm.)

\vskip .2 in 

For purposes of analysis, we might as well replace $\Sph2$ by a general compact space $K$.  Thus, let us consider
$$\align
M & = \Bbb R^1 \times (\alpha,\omega) \times K \\
\bar g & = -dt^2 + d\rho^2 + a(\rho)^2h_K
\endalign$$
where $h_K$ is a Riemannian metric on $K$.  We assume the conditions necessary on $a(\rho)$ so that $\hat M$ is formed as a cone over $[\alpha,\omega] \times K$, possibly squashed to a point at one end or the other.  This defines for us the topology on $\hat\partial(M)$: a pair of conjoined cones, each on either $K$ or a point.  

Let $\hat\partial_\iota(M)$ denote the cone on the end at $\iota$.  Let $i^+$ be the cone vertex; let $\hat M^0$ be the future completion with $i^+$ removed, and similarly $\hat\partial^0_\iota(M)$ the result of removing the vertex from either boundary cone, so that we have the decompositions $\hat M = \{i^+\} \cup  \hat M^0$, $\hat M^0 = M \cup \hat\partial^0(M)$, and $\hat\partial^0(M) = \hat\partial^0_\alpha(M) \cup \hat\partial^0_\omega(M)$.  We will want to focus on one end at a time, so it will help to also define $\hat M^0_\iota = M \cup \hat\partial^0_\iota(M)$.

For definiteness, let us consider the end at $\omega$; in particular, let us look at the interesting case, a cone on $K$.  So we will consider the possibilities for connections on $\hat\partial^0_\omega(M) = \R1 \times K$, attached to $M = \R1 \times (\alpha,\omega) \times K$ as $\fcb^0_\omega(\Minn)$ attaches to $(\widehat{\Minn})^0_\omega$.  Note that even to have a $C^0$-manifold structure we had to remove $i^+$; but with that removal, we have a clear $C^\infty$-manifold structure:  We just take the product with $K$ of the manifold with boundary $(\widehat{\Minn})^0_\omega$.

\head  2. Future Null Infinity for Static, Spherically Symmetric Spacetimes: Finding a Linear Connection 
\endhead

This section will be composed of an examination of why simple methods for extending the metric connection to this boundary fail; a brief review of affine connections on hypersurfaces; an application of such connections for finding limit connections on the causal boundary; and an examination of connections derived from conformal embeddings.

The question, then, is what possibilities are there for a ``natural" choice of connection on the smooth manifold $\R1 \times K$, smoothly embedded as $\{\sigma = \frac\pi2\}$ in $\{(\eta,\sigma) \st -\infty < \eta < \tan^{-1}(\tan\sigma - 2\alpha)\} \times K$?

\subhead Failure of simple methods for extending the metric connection
\endsubhead

If we have a manifold $N$, which is the interior of a manifold with boundary $\hat N$, what possibilities are there for extending a connection $\nabla$ on $N$ to $\partial(N)$?
  If the connection on $N$ is the Levi-Civita connection derived from a pseudo-Riemannian metric $g$ on $N$, then the obvious first question to ask is whether $g$ extends to a non-degenerate metric $g'$ (of any signature) on $\partial(N)$; if so, then the Levi-Civita connection $\nabla'$ for $g'$ provides the obviously natural connection, and $\nabla'$ is a natural extension of $\nabla$.  If $g$ extends to a degenerate metric $g'$ on $\partial(N)$, then we don't have a Levi-Civita connection from $g'$; but we can still inquire whether there is some torsion-free connection $\nabla'$ on $\partial(N)$ for which $\nabla' g' = 0$, and which also is in some sense an extension of $\nabla$.  If $g$ simply has no extension to $\partial(N)$, then we are left asking for some torsion-free extension of $\nabla$ to $\partial(N)$; and if that fails, there may be no simple answer.

The prototypical case to consider is for Minkowski space $\Lo^4$, which we can represent as $M = \R1 \times (0,\infty) \times \Sph2$ (removing a timelike line through the origin), co\"ordinates $(t, r, \phi, \theta)$ and metric $ds^2 = -dt^2 + dr^2 + r^2(d\phi^2 + \sin^2\phi\,d\theta^2)$.  We know the causal boundary for $\Lo^4$ is null, so we can't expect to use a Levi-Civita connection on the boundary.  But let us examine how far we get with trying to extend things from the spacetime to the boundary.

Define null co\"ordinates $\mu = t + r$ and $\lambda = t - r$, with range $\{\lambda < \mu\}$, yielding metric $ds^2 = -d\mu\,d\lambda + \frac14(\mu - \lambda)^2(d\phi^2 + \sin^2\phi\,d\theta^2)$.  Now define an associated null co\"ordinate $\sigma = \tan^{-1}\mu$, so we have range $\{(\lambda, \sigma)\,|\,\lambda < \tan\sigma \text{ and } |\sigma| < \frac\pi2\}$ and metric $ds^2 = -\sec^2\sigma\,d\sigma\,d\lambda + \frac14(\tan\sigma - \lambda)^2(d\phi^2 + \sin^2\phi\,d\theta^2)$.  It is this parametrization of $M$ which allows us to embed it in $\hat M^0_\infty$, which we render as $\{(\lambda, \sigma)\,|\,\lambda < \tan\sigma \text{ and } -\frac\pi2 < \sigma \le \frac\pi2 \} \times \Sph2$; then $\hat\partial^0_\infty(M)$ is rendered as $\sigma = \frac\pi2$, i.e., $\R1 \times \{\frac\pi2\} \times \Sph2$.  (The other end of the boundary is $\hat\partial^0_0(M)$, $\sigma = -\frac\pi2$ in $\hat M^0_0 = \{(\lambda, \sigma)\,|\,\lambda < \tan\sigma \text{ and } -\frac\pi2 \le \sigma < \frac\pi2 \} \times \Sph2$; this boundary element is timelike and a line, just replacing the timelike line removed from $\Lo^4$ to obtain $M$.)

We have $\hat M^0_\infty$ as a very nice manifold with boundary sitting inside $N = \R1 \times \R1 \times \Sph2$, which has global co\"ordinates $(\lambda, \sigma,\phi,\theta)$.  Let $\Lambda = \partial_\lambda$, $\Sigma = \partial_\sigma$, $\Phi = \partial_\phi$, and $\Theta = \partial_\theta$, the corresponding co\"ordinate tangent vectors in $N$.  (In terms of co\"ordinate vectors $T$ and $R$, derived from co\"ordinates $\{t, r\}$, we have $\Lambda = \frac12(T - R)$ and $\Sigma = \frac12(\sec^2)\sigma(T + R)$). In $M$ we have $|\Lambda| = |\Sigma| = 0$, $\<\Lambda,\Sigma\> = -\frac12\sec^2\sigma$, $|\Phi| = \frac12(\tan\sigma - \lambda)$, $|\Theta| = \frac12(\tan\sigma - \lambda)\sin\phi$, and other inner products 0.  The point to note here is that these vectors are defined on all of $N$, in particular on $\hat M^0_\infty$; but their inner products under the spacetime metric on $M$ do not extend to the boundary of $M$, $\hat\partial^0_\infty(M)$, in $N$:  Not only is $\hat\partial^0_\infty(M)$ null in the sense of causal boundary, but $g$ simply does not extend as even a continuous tensor from $M$ to $\hat\partial^0_\infty(M)$.  

Does the Levi-Civita connection $\nabla$ on $M$ extend to $\hat M^0_\infty$?  Let $U$ and $V$ be any tangent vector fields in $\Sph2$.  Then we have the following in $M$:
$$\align
\nabla_\Lambda\Lambda & = 0 \\
\nabla_\Lambda\Sigma & = 0 \\
\nabla_\Lambda V & = -\frac1{\tan\sigma - \lambda}V \\
\nabla_\Sigma \Sigma & = 2(\tan\sigma)\Sigma \\
\nabla_\Sigma V &  = \frac{\sec^2\sigma}{\tan\sigma - \lambda} V \\
\nabla_U V & = \frac12\<U,V\>_{\Sph2}(\tan\sigma - \lambda)(-(\cos^2\sigma)\Sigma + \Lambda) + \nabla^{\Sph2}_U V
\endalign$$
(where $\<-,-\>_{\Sph2}$ and $\nabla^{\Sph2}$ denote the standard Riemannian metric on the unit sphere and its corresponding Levi-Civita connection).  This plainly does not extend to $N$, as it blows up at $\sigma = \frac\pi2$ for $\nabla_\Sigma\Sigma$, $\nabla_\Sigma V$, and $\nabla_UV$ for any vector fields $U, V$ on $\Sph2$.  If all we are asking for is a connection on $\hat\partial^0_\infty(M)$, and not on all of $N$, then we need not be concerned about the first two of those, as $\Sigma$ is not a vector field on $\hat\partial^0_\infty(M) $.  But we rather need $\nabla_U V$ to be defined on the boundary, and what we have here is $\nabla_UV = -\frac12\<U,V\>_{\Sph2}rR + \nabla^{\Sph2}_U V$, with no limit on the boundary $\sigma = \frac\pi2$ (i.e., $r \to \infty$).

So we are left with trying some modification of the connection---hopefully in some ``natural" sense---that will give us a limit on the boundary.  Since we are interested in achieving a connection on a limiting hypersurface, perhaps it is best to start with hypersurface connections, submanifolds in the interior of $M$.  If we decide to look at timelike or spacelike hypersurfaces, there is the obvious choice of the induced Levi-Civita connection on each hypersurface; could that lead to a limit connection on our (null) boundary? 

Alas, this fails in the simplest of cases.  Consider as our null boundary $B$, the null cone in $\Lo^4$, $B = \{(x,y,z,t) \st t = \sqrt{x^2 + y^2 + z^2}\}$, and let $P_c$ be a more general cone, $P_c = \{(x,y,z,t) \st t = c\sqrt{x^2 + y^2 + z^2}\}$: $P_c$ is spacelike for $|c| < 1$, $P_c$ is timelike for $|c| > 1$, and $P_1 = B$.  Consider $P_c$ as $\{(\lambda p, c\lambda) \st \lambda > 0,\, p \in \Sph2\}$, so $P_c$ is parametrized on $(0,\infty)\times\Sph2$.  Let $\Lambda_c = \partial_\lambda$, the co\"ordinate vector in $P_c$.  For $c \neq 1$, we have the induced Levi-Civita connection on $\nabla^c$ on $P_c$.  It obeys 
$$\align
\nabla^c_{\Lambda_c}\Lambda_c & = 0 \\
\nabla^c_{\Lambda_c}V & = \frac1\lambda V \\
\nabla^c_UV & = \frac\lambda{-c^2 + 1}\<U,V\>_{\Sph2}\Lambda_c + \nabla^{\Sph2}_UV
\endalign$$
for vector fields $U$ and $V$ in $\Sph2$.  But there is no limit of $\nabla^c_U V$ as $c \to 1$.   Something very similar happens if we try approximating $B$ by quadric surfaces.

So a purely metric approach to finding a limit connection on the boundary seems stymied---perhaps to be expected in the case of a null boundary.  But we can approach the problem from a non-metric perspective.

\subhead Affine connections for hypersurfaces with transversals
\endsubhead

The approach to be followed here will be to look for connections on our boundary $B$ derived from connections $\nabla^n$ on hypersurfaces $\{P_n\}$ in $M$ approaching $B$; but with $B$ typically being null, we will take the hypersurfaces also to be null, which means we do not have access to a Levi-Civita connection.  Instead we will consider affine connections on our hypersufaces, as induced by choices of a transverse vector field; a modern reference for this is \cite{NSs}, sections II.1 and II.2 (much of the development of this theory is due to Blaschke in the early twentieth century).

General definitions: If $M$ is a manifold and $P$ an embedded hypersurface, then by a {\it transverse}\/ vector field $\xi$ on $P$, let us mean, for each $x \in P$, a choice of $\xi_x \in T_xM$ with $\xi_x \notin T_xP$, so that $T_xP + \text{span}(\xi_x) = T_xM$.  If $M$ comes equipped with an affine (i.e., linear) connection $\nabla$, then we have the {\it induced affine connection}\/ $\nabla^\xi$ on $P$, with accompanying {\it affine fundamental form}\/ $H$ on $P$ (a 2-covariant tensor) defined by
$$\Del_XY = \Del^\xi_XY + H(X,Y)\xi$$
for all $P$-vector fields $X$ and $Y$ on $P$; if $\Del$ is torsion-free, then so is $\Del^\xi$ and $H$ is symmetric.  We also obtain the {\it shape operator}\/ $S$ (a (1,1)-tensor) and the {\it transversal connection form} $\tau$ (a 1-form) on $P$, defined by
$$\Del_X\xi = -S X + \tau(X)\xi$$
for any $X \in T_xP$. 

There is, of course, a great deal of freedom in choosing the transverse vector field $\xi$.  Proposition 2.5 of \cite{NSs} summarizes the changes in the connection and the various tensors in changing from $\xi$ to $\bar\xi$, where $\bar\xi = \phi\xi + Z$ for choice of scalar function $\phi$ and vector field $Z$ on $P$:
$$\align
\bar H & = (1/\phi)H \\
\Del^{\bar\xi}_XY & = \Del^\xi_XY - \bar H(X,Y)Z \\
\bar\tau & = \tau + \bar H(-,Z) + d\ln|\phi| \\
\bar S & = \phi S - \Del^\xi_{(-)}Z + \bar\tau\otimes Z
\endalign$$
In case $M$ has a $\nabla$-parallel volume form $\eta$, then the induced volume form $\eta_\xi = \eta(\xi, \dots)$ on $P$ is parallel in $\nabla^\xi$ iff $\tau = 0$ (Proposition II.1.4 of \cite{NSs}); when this happens, $\xi$ is called an {\it equiaffine}\/ transversal vector field.

\subhead Limiting affine connections for Future Null Infinity
\endsubhead

Once again considering our standard context:  $M = \R1 \times (\alpha,\omega) \times K$, $\hat M^0_\omega = (\widehat{\Minn})^0_\omega \times K$, $\fcb^0_\omega(M) = \fcb^0_\omega(\Minn)\times K$; $K$ is compact with Riemannian metric $h_K$, and $M$ has spacetime metric $ds^2 = F(\rho)(-dt^2 + d\rho^2 + a(\rho)^2 h_K$) (connection with previous notation:  $F(\rho) = \Omega(r)$).  We assume $\int_{\rho_0}^\omega \frac1{a(\rho)^2}\, d\rho < \infty$ and that $a(\rho)$ is eventually increasing for $\rho \to \omega$ (so that the causal boundary component at $\omega$ is a cone on $K$, justifying the description of $\hat M^0_\omega$).  Specializing to $\omega = \infty$ gives us a null boundary.

We define null co\"ordinates $\lambda
 = t - \rho$, $\sigma = \tan^{-1}(t + \rho)$.  These (along with co\"ordinates on $K$) give a parametrization of $M$ with $\tan\sigma - 2\omega < \lambda < \tan\sigma - 2\alpha$ (and $\sigma$ takes on all values in $(-{\frac\pi2},{\frac\pi2})$, as $t$ takes on all values in $\R1$); typically, $(\alpha,\omega) = (0, \infty)$, and we just have $\lambda < \tan\sigma$.  Let $\Lambda = \partial_\lambda$ and $\Sigma = \partial_\sigma$ be the co\"ordinate vectors; they are null.  For any $\sigma_0 \in ({\frac\pi2},{\frac\pi2})$, let $B^{\sigma_0}$ be the $\sigma = \sigma_0$ slice in $M$; this is a null hypersurface parametrized as $\{(\lambda,p) \in (\tan\sigma_0 - 2\omega, \tan\sigma_0 - 2\alpha) \times K\}$; and with $\omega = \infty$, any $(\lambda,p)$ will eventually be included in $B^{\sigma_0}$ as $\sigma_0 \to {\frac\pi2}$.  We will examine the affine connection $\nabla^\xi$ induced on $B^{\sigma_0}$ by a transverse vector field $\xi$ and the Levi-Civita connection $\nabla$ on $M$.

Considering that our actual motivation is for a spherically symmetric spacetime, it is only sensible to restrict $\xi$ to spherically symmetric behavior:  We take $\xi$ to lie in span$\{\Sigma, \Lambda\}$, so $\xi = \psi\Sigma + \beta\Lambda$ for scalar functions $\psi$ and $\beta$ with no spherical dependence---or, in our more general setting, no $K$-dependence.  We are, in fact, considering these slices $B^\sigma$ for all $\sigma$, so we have $\xi = \psi(\sigma,\lambda)\Sigma + \beta(\sigma,\lambda)\Lambda$.  The only absolute requirement on $\xi$ is that $\psi$ is never 0.  Then we have the following (where $U$ and $V$ are co\"ordinate vector fields in $K$ and $\nabla^K$ is the Levi-Civita connection in $K$) on $B^\sigma$:
$$\align
\nabla^\xi_\Lambda\Lambda & = -\frac{F'}{2F}\Lambda \\
\nabla^\xi_\Lambda V & = -\frac12\(\frac{F'}{2F} + \frac{a'}a\)V \tag 1\\
\nabla^\xi_U V & = a^2\(\frac{F'}{2F} + \frac{a'}a\)\(\frac\beta\psi\cos^2\sigma + 1\)h_K(U,V)\Lambda + \nabla^K_U V 
\endalign$$
with the associated tensors
$$\align
H(\Lambda,\Lambda) & = 0 \\
H(\Lambda,V) & = 0 \\
H(U,V) & = -a^2\(\frac{F'}{2F} + \frac{a'}a\)\frac1\psi\(\cos^2\sigma\)h_K(U,V) \\
S\Lambda & = \beta\(\frac{\Lambda\psi}\psi - \frac{\Lambda\beta}\beta + \frac{F'}{2F}\)\Lambda \\
SU & = \frac12\(\frac{F'}{2F} + \frac{a'}a\)\(\beta - \psi\sec^2\sigma\)U \\
\tau(\Lambda) & = \frac{\Lambda\psi}\psi \\
\tau(U) & = 0
\endalign$$
In particular, we can easily impose the condition that the volume form on $B^\sigma$, induced via $\xi$ from the metric volume-form on $M$, be $\nabla^\xi$-parallel (i.e., that $\xi$ be equiaffine):  We just require that $\Lambda\psi = 0$, i.e., that $\psi = \psi(\sigma)$, no $\lambda$-dependence.  With this assumption, we have
$$\align
H & = -a^2\frac12\(\ln(F a^2)\)'\frac{\cos^2\sigma}\psi h_K \\
S & = \beta\(-\Lambda(\ln|\beta|) + \frac12(\ln F)'\)d\lambda\otimes\Lambda \\
& \quad + \frac12\(\ln(F a^2)\)'\frac\psi{\cos^2\sigma}\(\frac\beta\psi\cos^2\sigma - 1\)\text{id}_K
\endalign$$
This suggests a further constraint on $\xi$, that $\beta$ also have no $\lambda$-dependence, simplifying the first term for $S$---or perhaps the stronger condition that we simplify $S$ as much possible by insisting $\beta = \psi \sec^2\sigma$, yielding $S = \frac12(\ln F)'(\psi\sec^2\sigma)d\lambda \otimes \Lambda$ and $\nabla^\xi_U V = a^2\(\ln(Fa^2)\)'h_K(U,V) + \nabla^K_UV$.

The object now is to reach a connection $\hat\nabla$ on $B = \hat\partial^0_\omega(M)$ by taking the limit of the connections $\nabla^\xi$ on $B^\sigma$ as $\sigma \to {\frac\pi2}$.  Since the co\"ordinates we're using extend to $B$, so do the co\"ordinate vector fields; so all we need is limits for the expressions (1), that is to say, we require there exist (as finite limits)
$$\align
p(\lambda) & = \lim_{\sigma \to {\frac\pi2}} \(\ln(F(\rho)\)' \\
q(\lambda) & = \lim_{\sigma \to {\frac\pi2}} \(\ln(F(\rho)a(\rho)^2)\)' \tag 2 \\
w(\lambda) & = \lim_{\sigma \to {\frac\pi2}} \frac12a(\rho)^2\(\ln(F(\rho)a(\rho)^2)\)'
\(\frac{\beta(\lambda,\sigma)}{\psi(\lambda,\sigma)}\cos^2\sigma + 1\)
\endalign$$
where $\rho = \frac12(-\lambda + \tan\sigma)$ (and the prime indicates $\frac d{d\rho}$); the first two conditions are requirements on the spacetime metric, but the third is a requirement on our choice of transversal vector field $\xi$.  Actually, as $\sigma \to \frac\pi2$, we have $\rho \to \infty$, so $p$ and $q$ in (2) must be constants, not functions of $\lambda$; but it is possible for $w(\lambda)$ to actually depend on $\lambda$ and still have a limiting connection exist.  Assuming the existence of these limits, we then obtain a connection on $B$:
$$\align
\hat\nabla_\Lambda\Lambda & = -\frac12p\Lambda \\
\hat\nabla_\Lambda V & = -\frac14qV \tag 3\\
\hat\nabla_U V & = w(\lambda)h_K(U,V)\Lambda + \nabla^K_U V
\endalign$$

Then the question becomes, is $\hat\nabla$ geodesically complete?  And does the answer to that depend on the choice we make of transversal fields $\xi$---that is to say, on $w(\lambda)$?  For while $p$ and $q$ are inherent in the spacetime, $w$ depends on the extraneous field $\xi$.

First let us note that, irrespective of $\xi$, $\hat\nabla$ cannot be compete unless $p = 0$:  The integral curves of $\Lambda$ are pregeodesics in $B$, and a geodesic along such a path has the form $c(s) = (\lambda(s),x_0)$ for some $x_0 \in K$ (parametrizing $B$ as $\R1\times K$), with $\lambda(s)$ satisfying   $\lambda'' - \frac12p\lambda'{}^2 = 0$; for $p \neq 0$, this has solution $\lambda(s) = \lambda_0 - \frac2p\ln\(1 - \frac p2\lambda'_0(s - s_0)\)$ (where $\lambda_0 = \lambda(s_0)$, $\lambda'_0 = \lambda'(s_0)$), which is not defined for all real $s$.

Whether $q$ being non-zero prevents completeness depends on the behavior of $w(\lambda)$.  For a curve of the form $c(s) = (\lambda(s),x(s))$, the geodesic equation (assuming $p = 0$) gives us $\lambda'' + w(\lambda)h_K(\dot x(s),\dot x(s)) = 0$.  Let $v_0 = h_K(\dot x(0), \dot x(0))$, which we take to be non-zero; then the solution is given by
$$s - s_0 = \pm\int_{\lambda_0}^{\lambda(s)} \frac{d\mu}{\sqrt{\lambda'_0{}^2 - 2v_0^2e^{-q\lambda_0}\int_{\lambda_0}^{\mu} w(u)e^{qu}du }}$$
Let's look at what happens with the simplest assumption about $w$---that it's constant.  Then if $q \neq 0$ we get
$$s - s_0 = \pm\int_{\lambda_0}^{\lambda(s)} \frac{d\mu}
{\sqrt{\lambda'_0{}^2 + 2v_0^2\frac{w}q\(1 - e^{q(\mu - \lambda_0)}\)}}$$
which will be complete if and only if the integral is infinite for $\lambda(s)$ going to its maximum and minimum values.  First suppose $\frac wq < 0$; then for large values of $|\mu|$ with $q\mu > 0$, the integrand behaves as $\frac1{v_0}e^{-(q/2)(\mu - \lambda_0)}$, which has a finite integral as $\lambda(s)$ goes to the appropriate infinity.  Now suppose $\frac wq > 0$; this puts a bound on allowed values of $\mu$ (i.e., either the min or max value for $\lambda$ is finite).  The quantity under the radical, as $\mu$ approaches this limiting value for $\lambda$, behaves just like $C - e^x$ as $x \to (\ln C)^-$, i.e., like $C(1 - e^{x - \ln C}) = C(-(x - \ln C) - \frac12(x - \ln C)^2 - \cdots)$ as $x - \ln C \to 0^-$; so the integral behaves like $\int_{y_0}^0 y^{-\frac12} dy$---again, finite.  Thus, $q \neq 0$ produces an incomplete limit connection on the causal boundary, with the assumption that $w$ is constant.

However, standard cosmological models such as Minkowski space, Schwarzschild, and Reissner-Nordstr\"om, have $p = q = 0$.  Let us gather together the common behaviors found in these models and call them the Cosmological Conditions:

\definition{Cosmological Conditions} We will say a static, spherically symmetric spacetime $(M,g) = (\R1 \times (\alpha,\omega) \times \Sph2,\, F(\rho)(-dt^2 + d\rho^2 + a(\rho)^2 h_{\Sph2}))$ (or, more generally, a static spacetime $(\R1 \times (\alpha,\omega) \times K,\, F(\rho)(-dt^2 + d\rho^2 + a(\rho)^2 h_K))$ for any compact Riemannian manifold $(K,h_k)$) satisfies {\it the Cosmological Conditions}\/ at $\omega$ if
\roster
\item $\lim_{\rho \to \omega} (\ln(F(\rho))' = 0$ 
\item $\lim_{\rho \to \omega} \(\ln(F(\rho)a(\rho)^2)\)' = 0$
\item $\int_{\rho_0}^\omega \frac1{a(\rho)^2} d\rho < \infty$ \quad (for some $\rho_0 \in (\alpha,\omega)$) 
\item $a(\rho)$ is eventually increasing as $\rho \to \omega$  
\endroster
(Alternatively: Conditions (1) and (2) are the same as saying $F'/F$ and $a'/a$ both have limit 0 as $\rho \to \omega$.)
\enddefinition

In the Common Model Spacetime Paradigm, i.e., $ds^2 = -f(r)dt^2 + \frac1{f(r)}dr^2 + r^2h_{\Sph2}$ on $\R1 \times (r_{\text{min}},r_{\text{max}}) \times \Sph 2$ with either $r_{\text{min}} > 0$ and $\frac{r^2}{f(r)}$ eventually increasing towards either end of the interval, or with $r_{\text{min}} = 0$ and $\frac{r^2}{f(r)}$ eventually increasing as $r \to r_{\text{max}}$ but eventually decreasing to 0 as $r \to 0$, the Cosmological Conditions at the $r_{\text{max}}$ end of the interval amount to 
\roster
\item $\lim_{r \to r_{\text{max}}} f'(r) = 0$ 
\item $\lim_{r \to r_{\text{max}}} \frac1{f(r)} = 0$
\endroster
as Cosmological Conditions (3) and (4) are already satisfied.  (Full translation from the Common Model Spacetime Paradigm to the standard static presentation is $F(\rho) = f(r)$, $a(\rho) = \frac{r}{\sqrt{f(r)}}$,  $d\rho = \frac{dr}{f(r)}$, and $(\alpha,\omega) = (r_{\text{min}},r_{\text{max}})$.) We may note in particular that Minkowski space, Schwarzschild, and Reissner-Nordstr\"om satisfy the Cosmological Conditions at infinity, but Schwarzschild--Anti-de Sitter does not (nor, {\it a fortiori\/}, does Schwarzschild-de Sitter).

The Cosmological Conditions having been defined, we can now state the main result of this section.

\proclaim{Theorem 2.1} Let $M$ be a standard static spacetime of the form $M = \R1 \times (\alpha,\infty) \times K$ for $K$ a compact manifold, $M$ having metric $ds^2 = F(\rho)(-dt^2 + d\rho^2
 + a(\rho)^2 h_K)$ (where $h_K$ is a Riemannian metric on $K$), satisfying the Cosmological Conditions at $\infty$.  Then the future causal boundary component at $\infty$, $\hat\partial^0_\infty(M)$, has the smooth manifold structure of $\R1\times K$, attached to $M$ in the manner of manifold with boundary, and the causal structure on $M$ extends to $\hat M^0_\infty = M \cup \hat\partial^0_\infty(M)$ to make the boundary a null hypersurface.  

Let $\lambda = -\rho + t$ and $\sigma = \tan^{-1}(\rho + t)$ on $M$, and for any $\sigma_0 \in (-{\frac\pi2},{\frac\pi2})$, let $B^{\sigma_0}$ be the null hypersurface $\sigma = \sigma_0$. Then for any smooth function $w: \R1 \to \R1$, there is a connection $\hat\nabla$ on $\hat\partial^0_\infty(M)$ such that, in terms of co\"ordinate vectors derived from $\lambda$ and co\"ordinates on $K$, $\hat\nabla$ is the limit as $\sigma \to {\frac\pi2}$ of equiaffine connections on each $B^\sigma$ and obeys (for $\Lambda = \partial_\lambda$, $U$ and $V$ co\"ordinate vectors in $K$, and $\nabla^K$ the Levi-Civita connection in $K$),
$$\align
\hat\nabla_\Lambda\Lambda & = 0 \\
\hat\nabla_\Lambda V & = 0 \\
\hat\nabla_UV & = w(\lambda)h_K(U,V)\Lambda + \nabla^K_UV
\endalign$$
This is the only possible form for a connection obtained in this manner (limit of equiaffine connections on $B^\sigma$).

For $w$ chosen to be a constant function, $\hat\nabla$ is geodesically complete.  (More generally, $\hat\nabla$ is complete precisely when the function $\lambda \mapsto \int_{\lambda_0}^\lambda d\mu/\sqrt{1 - \int_{\lambda_0}^\mu w(u)\,du}$ attains arbitrarily large absolute values in at least one direction.)

The connection $\hat\nabla$ has all the symmetries that $K$ does; in particular, if $K$ is a round sphere, $\hat\nabla$ has spherical symmetry. 
\endproclaim

\demo{Proof}  We already know the topological and causal nature of $\hat\partial^0_\infty(M)$.  Equations (2) and (3) show us how to select the transversal field $\xi$ in each $B^\sigma$:  For instance, choose $\psi(\lambda,\sigma) = 1$  and $\beta(\lambda,\sigma) =  \(2w(\lambda)/\(a(\rho)^2\(\ln(F(\rho)a(\rho)^2)\)'\) - 1\)\sec^2(\sigma)$.  (In particular, we can arrange $w = 0$ by selecting $\beta = -\psi\sec^2(\sigma)$.)  

Note the Cosmological Conditions and equations (2) require $\hat\nabla_\Lambda\Lambda = 0$ and $\hat\nabla_\Lambda V = 0$.

The geodesic equation under $\hat\nabla$ for the curve $(\lambda(s),x(s))$ in $\R1\times K$ is
$$\align
\lambda'' & = -w(\lambda)h_K(\dot x,\dot x) \\
\nabla^K_{\dot x}\dot x & = 0
\endalign$$ 
so the curve $x(s)$ is just a geodesic in the compact Riemannian manifold $K$; it is necessarily complete (defined for all $s \in \R1$), and it has $h_K(\dot x,\dot x)$ constant.  For $w(\lambda)$ a non-zero constant, $\lambda(s)$ has the obvious quadratic solution producing geodesics which reach a maximum (or minimum) value for $\lambda$ before turning around and heading back to $\lambda \to -\infty$ (or $\infty$); for $w = 0$, $\lambda$ is just linear in $s$ (and we have the product connection on $\R1 \times K$).   More generally, $\lambda(s)$ has the solution 
$$\pm(s - s_0) = \int_{\lambda_0}^{\lambda(s)} \frac{d\mu}{\sqrt{\lambda'_0{}^2 - 2k^2 \int_{\lambda_0}^\mu w(u)du}}$$
where $k^2 = h_K(\dot x_0,\dot x_0)$. (In particular, for $w(\lambda)$ being a non-zero constant, the geodesic turns around when $\lambda(s) = \lambda_0 + \lambda'_0{}^2/(2k^2w).)$ Whether or not the integral will reach arbitrary values of $s$ is independent of the values for $\lambda'_0$ and $k$. \qed
\enddemo

\subhead Relationship with conformal boundary
\endsubhead

The complex of $\hat\nabla$ connections from Theorem 1.1---parametrized by functions $w(\lambda)$---is an arguably natural collection, in that it comes from a limit of equiaffine connections (spherically symmetric, when $K$ is a sphere) on naturally selected null hypersurfaces.  Among this complex, the clearly simplest are with $w$ being constant, and they are all complete (and the $w = 0$ connection is the product connection on $\R1 \times K$).

We should ask how this complex of connections compares with that from looking at conformal boundaries on $M$.  More precisely, let us assume that the topological boundary we need to examine is the future causal boundary as above, so that we will be concerned with the obvious embedding of $M = \R1 \times (\alpha,\infty) \times K$ into $\bar M$ by extension of the null co\"ordinate $\sigma$ (defined as before) from $(-{\frac\pi2},{\frac\pi2})$ to $(-{\frac\pi2},{\frac\pi2} + \epsilon)$.  (By Theorems 4.16 and 4.26 in \cite{FHeSn2}, a conformal boundary which is one-sided and $C^1$ except for isolated points, like $i^+$, that are well-behaved, will have the same topological and causal characteristics as the future causal boundary.)  What this amounts to is selecting a positive scalar function $G$ on $M$ so that $\bar g = Gg$ extends smoothly to a metric on $\bar M$.  To insure spherical symmetry (in the case $K = \Sph 2$), we will insist that $G = G(\lambda,\sigma)$, i.e., $G$ is independent of the $K$-factor; what we need is that $\bar g = G(\lambda,\sigma)g$ extends as a smooth metric past $\sigma = {\frac\pi2}$.  Then $(\bar M, \bar g)$ has a Levi-Civita connection $\bar\nabla$.  The boundary $\hat\partial^0_\infty(M)$ in $\bar M$, i.e., $\{\sigma = {\frac\pi2}\}$, is null, so there is no induced connection on the boundary---unless the boundary is totally geodesic in $\bar M$, in which case we have $\hat{\bar\nabla}$, the restriction of $\bar\nabla$ to the boundary.  This connection is in no way a limit of $\nabla$ on $M$, but it may perhaps be considered a reasonable connection on the boundary; of course, it is a complex of connections, depending on the function $G$.

What is interesting is that it is the same constellation of connections as given in Theorem 2.1; but it applies only in a restricted class of spacetimes, those for which $a'(\rho)$ has a finite, non-zero limit as $\rho \to \infty$.

\proclaim{Theorem 2.2} Let $M$ be a standard static spacetime of the form $M = \R1 \times (\alpha,\infty) \times K$ for $K$ a compact manifold, $M$ having metric $ds^2 = F(\rho)(-dt^2 + d\rho^2 + a(\rho)^2 h_K)$ (where $h_K$ is a Riemannian metric on $K$), satisfying the Cosmological Conditions at $\infty$.  With $M$ given the null co\"ordinates $\lambda = \frac12(-t + \rho)$ and $\sigma = \tan^{-1}(\frac12(t + \rho))$, consider $M$ as embedded in $\bar M$ by extension of $\sigma$ and with future causal boundary component $\hat\partial^0_\infty(M)$ occurring as $\{\sigma = {\frac\pi2}\}$.

Suppose that $\lim_{\rho \to \infty} a'(\rho)$ exists as finite and non-zero.  Consider Lorentz metrics $\bar g$ on $\bar M$ such that (1) for some positive $K$-independent scalar function $G: M \to \Bbb R^1$, $\bar g = Gg$ on that portion of $\bar M$ identifiable with $M$ in the obvious way, and (2) $\hat\partial^0_\infty(M)$, as a hypersurface in $\bar M$, is totally geodesic with respect to $\bar g$.  Then the connections that arise on $\hat\partial^0_\infty(M)$ as the restriction of the Levi-Civita connection on $(\bar M,\bar g)$ are precisely the connections from Theorem 2 .1.

If, on the other hand, $\lim_{\rho \to \infty} a'(\rho)$ is zero, infinite, or non-existent, then there is no metric $\bar g$ on $\bar M$ such that the embedding of $(M, g)$ into $(\bar M, \bar g)$ is conformal with a $K$-independent conformal factor.
\endproclaim

\demo{Proof} For a given function $G(\lambda,\sigma)$, we have the following values for the metric $\bar g = Gg$ (where, as usual, $U$ and $V$ are vectors in $K$):
$$\align
\bar g(\Lambda,\Lambda) & = 0 \\
\bar g(\Lambda,\Sigma) & = G(\lambda,\sigma)(-\tfrac12)F(\rho)\sec^2(\sigma) \\
\bar g(\Lambda, V) & = 0 \tag 4\\
\bar g(\Sigma,\Sigma) & = 0 \\
\bar g(\Sigma, V) & = 0 \\
\bar g(U,V) & = G(\lambda,\sigma)F(\rho)a(\rho)^2h_K(U,V)
\endalign$$
so that in order to have $Gg$ form a metric $\bar g$ on the boundary, at $\sigma = {\frac\pi2}$, we must have the following finite limits exist and be non-zero:
$$\align
\lim_{\sigma \to {\frac\pi2}} G(\lambda,\sigma)F(\rho)\sec^2\sigma & = A(\lambda) \\
\lim_{\sigma \to {\frac\pi2}} G(\lambda,\sigma)F(\rho)a(\rho)^2 & = B(\lambda) \\
\endalign$$
In particular, note that we have $B(\lambda)/A(\lambda) = \lim_{\sigma \to {\frac\pi2}} (a(\rho)\cos\sigma)^2$.  It follows that $\lim_{\sigma \to {\frac\pi2}} a(\rho)\cos\sigma$ must exist, be finite, and be non-zero.  This implies $a(\rho)$ must grow to infinity as $\rho \to \infty$, for otherwise that limit would be zero (recall that $a(\rho)$ is eventually increasing as $\rho \to \infty$).  Then by l'H\^opital, we have  
$$\align 
\lim_{\sigma \to \frac\pi2} a(\rho)\cos(\sigma) & = 
\lim_{\sigma \to \frac\pi2} \frac{\frac{da(\rho)}{d\sigma}}{\sec\sigma\tan\sigma} \\ 
& = \lim_{\sigma \to \frac\pi2} \frac{a'(\rho)\frac12\sec^2\sigma}{\sec\sigma\tan\sigma} 
\\
& = \tfrac12\lim_{\sigma \to \frac\pi2} \frac{a'(\rho)}{\sin\sigma} \\
& = \tfrac12\lim_{\rho \to \infty} a'(\rho)
\endalign$$
(For convenience, we might as well write as $\lim_{\sigma \to {\frac\pi2}} a(\rho)\cos\sigma = \frac12a'(\infty)$.)  

So we have established the last paragraph in the theorem.  For the balance, we assume  $a'(\infty)$ is finite and non-zero (implying $a(\infty) = \infty$).

From equations (4) we have
$$\align
\bar\nabla_\Lambda\Lambda & = \(\frac{G_\lambda(\lambda,\sigma)}{G(\lambda,\sigma)} - \frac{F'(\rho)}{2F(\rho)}\)\Lambda \\
\bar\nabla_\Lambda V & = \frac12\(\frac{G_\lambda(\lambda,\sigma)}{G(\lambda,\sigma)} - \frac{F'(\rho)}{2F(\rho)} - \frac{a'(\rho)}{a(\rho)}\)V \\
\bar\nabla_U V & = h_K(U,V)a(\rho)^2\cos^2\sigma\(\(\frac{G_\lambda(\lambda,\sigma)}{G(\lambda,\sigma)} - \frac{F'(\rho)}{2F(\rho)} - \frac{a'(\rho)}{a(\rho)}\)\Sigma \right. \\
& \quad\quad\quad\quad\quad + \left.\(\frac{G_\sigma(\lambda,\sigma)}{G(\lambda,\sigma)}
 + \(\frac{F'(\rho)}{2F(\rho)} + \frac{a'(\rho)}{a(\rho)}\)\sec^2\sigma\)\Lambda\) + \nabla^K_U V
\endalign$$
Using the Cosmological Conditions, we have the following for limits as $\sigma \to {\frac\pi2}$:
$$\align
\bar\nabla_\Lambda\Lambda & = \(\lim_{\sigma \to \frac\pi2}\frac{G_\lambda(\lambda,\sigma)}{G(\lambda,\sigma)}\)\Lambda \\
\bar\nabla_\Lambda V & = \frac12\(\lim_{\sigma \to \frac\pi2}\frac{G_\lambda(\lambda,\sigma)}{G(\lambda,\sigma)}\)V \\
\bar\nabla_U V & = h_K(U,V)\frac14a'(\infty)^2\(\(\lim_{\sigma \to \frac\pi2}\frac{G_\lambda(\lambda,\sigma)}{G(\lambda,\sigma)} \)\Sigma \right. \\
& \quad\quad\quad\quad\quad + \left.\lim_{\sigma \to \frac\pi2}\(\frac{G_\sigma(\lambda,\sigma)}{G(\lambda,\sigma)} + \(\frac{F'(\rho)}{2F(\rho)} + \frac{a'(\rho)}{a(\rho)}\)\sec^2\sigma\)\Lambda\) + \nabla^K_U V
\endalign$$
Thus we see that the boundary $\{\sigma = {\frac\pi2}\}$ is totally geodesic in $\bar M$---and, thus, yields a connection $\hat{\bar\nabla}$ on the boundary through restriction to vector fields lying in the boundary---iff  $\lim_{\sigma \to \frac\pi2}\frac{G_\lambda(\lambda,\sigma)}{G(\lambda,\sigma)} = 0$.  Assuming that to be the case, we have, for the connection on the boundary,
$$\align
\hat{\bar\nabla}_\Lambda\Lambda & = 0 \\
\hat{\bar\nabla}_\Lambda V & = 0 \\
\hat{\bar\nabla}_U V & = h_K(U,V)\frac14a'(\infty)^2\lim_{\sigma \to \frac\pi2}\(\frac{G_\sigma(\lambda,\sigma)}{G(\lambda,\sigma)} + \(\frac{F'(\rho)}{2F(\rho)} + \frac{a'(\rho)}{a(\rho)}\)\sec^2\sigma\)\Lambda + \nabla^K_U V
\endalign$$

We've almost achieved the same collection of connections on the boundary as from Theorem 1.1; all that remains is to demonstrate we can have that last limit be any $w(\lambda)$ while maintaining $\lim_{\sigma \to \frac\pi2}\frac{G_\lambda(\lambda,\sigma)}{G(\lambda,\sigma)} = 0$.  But this is easily done:  For any smooth function $w(\lambda)$, define
$$G^w(\lambda,\sigma) = \frac1{F(\rho)a(\rho)^2}e^{\(\frac4{a'(\infty)^2}w(\lambda)(\sigma - \frac\pi2)\)}$$  
Note that $G^w_\lambda/G^w = \frac4{a'(\infty)^2}w'(\lambda)(\sigma - \frac\pi2)$, vanishing at $\sigma = {\frac\pi2}$.  Then using $G^w$ for $G$ yields $\bar g = \exp{\(\frac4{a'(\infty)^2}w(\lambda)(\sigma - \frac\pi2)\)}\(-\frac12\frac{d\lambda\,d\sigma}{(a(\rho)\cos\sigma)^2}  + h_K\)$ and
$$\align
\bar\nabla_\Lambda\Lambda & = \frac4{a'(\infty)^2}w'(\lambda)\(\sigma - \frac\pi2\)\Lambda \\
\bar\nabla_\Lambda V & = \frac2{a'(\infty)^2}w'(\lambda)\(\sigma - \frac\pi2\)V \\
\bar\nabla_U V & = h_K(U,V)w(\lambda)\Lambda + \nabla^K_U V
\endalign$$
so at the boundary we have
$$\align
\hat{\bar\nabla}_\Lambda\Lambda & = 0 \\
\hat{\bar\nabla}_\Lambda V & = 0 \\
\hat{\bar\nabla}_U V & = h_K(U,V)w(\lambda)\Lambda + \nabla^K_U V
\endalign$$
as desired.
\qed
\enddemo

For the common model spacetime paradigms, we have $a'(\rho) = \frac{f(r) - \frac12rf'(r)}{\sqrt{f(r)}}$.  For Minkowski space, Schwarzschild, and Reissner-Nordstr\"om, this gives $a'(\infty) = 1$.  (Schwarzschild--Anti-de Sitter gives $a'(\infty) = 0$; but this doesn't satisfy the Cosmological Conditions, anyway.)

\head 3. Other Boundaries: Null, Timelike, and Spacelike
\endhead

Let us consider other boundaries that can occur for this class of spacetimes, i.e., $F(\rho)(-dt^2 + d\rho^2 + a(\rho)^2h_K)$ on $\R1 \times (\alpha,\omega) \times K$ (for $K$ compact) or their natural extensions.  What sort of connections come naturally?  In brief:  It's much simpler than for a null $\R1\times K$.

\subhead Christodoulou-completion at $\Cal I^+$
\endsubhead

Christodoulou in \cite{C} (see also Dafermos, \cite{D}) considers a sort of completeness at $\Cal I^+$---meaning null future boundary in a spherically symmetric, asymptotically flat spacetime---for purposes of a kind of weak cosmic censorship.  

Although the actual causal (or conformal) boundary in this conception has the topology and causality of a null cone on $\Sph2$, the only question raised about completion in this treatment lies along the null line factor.  For the static case, then, it's a comparatively trivial matter.  

The actual definition of completeness of $\Cal I^+$ in \cite{C}, applied to our context here, is this:  For any $x \in \Sph2$ and any $\sigma < {\frac\pi2}$, consider the future null geodesic $\gamma_{\sigma_0,x}(s) = (\lambda(s),\sigma_0,x)$ with $\lambda(0) = 0$ and $\lambda'(0) = 1/\sqrt{F(\frac12\tan\sigma_0)}$; then it is required that the maximum domain $[0,s^{\sigma_0}_{\text{max}})$ for $\gamma_{\sigma_0,x}$ (clearly independent of $x \in \Sph2$) grow to $[0, \infty)$ as $\sigma_0 \to {\frac\pi2}$. 

The geodesic equation for $\gamma_{\sigma_0,x}$ amounts to $\lambda''(s) = \frac{F'(\rho)}{2F(\rho)}\lambda'(s)$ for $\rho = \frac12(-\lambda + \tan\sigma_0)$, giving a solution of $\sqrt{F(\tan\sigma_0)}(s - s_1) = \int_{-\frac12(\lambda(s) + \tan\sigma_0)}^{\tan\sigma_0} F(u)\, du$ , with $s_1$ defined by $\lambda(s_1) = -\tan\sigma_0$; so we obtain $s^{\sigma_0}_{\text{max}} = \frac1{\sqrt{F(\tan\sigma_0)}}\int_\alpha^{-\frac12(\lambda_0 + \tan\sigma_0)} F(u)\,du$.  Then $s_{\text{max}} = \lim_{\sigma_0 \to {\frac\pi2}} s^{\sigma_0}_{\text{max}} = \lim_{\rho\to\infty} \frac1{\sqrt{F(\rho)}} \int_\alpha^\rho F(u)\,du$.  With the cosmological assumption $\frac{F'}F \to \infty$, this always results in $s_{\text{max}} = \infty$. (Proof: For otherwise, for some constant $A$, we have $\int_{\rho_0}^\rho F(u)\,du < A\sqrt{F(\rho)}$ for all $\rho$, i.e., for $\Phi(\rho) = \int_{\rho_0}^\rho F(u)\,du$, $\frac1{A^2} < \frac{\Phi'}{\Phi^2}$.  Pick some $\rho_1 > \rho_0$; then $\frac1{A^2}(\rho - \rho_1) < \frac1{\Phi(\rho_1)} - \frac1{\Phi(\rho)}$.  But that is false for $\rho \ge \rho_1 + \frac{A^2}{\Phi(\rho_1)}$.)

Thus we have the result:

{\it A static spherically symmetric spacetime $M$ with $\omega = \infty$ and obeying the Cosmological Conditions at $\infty$ with null $\fcb^0_\infty(M)$ is Christodoulou-complete}.  

This is not precisely the same as defining a complete connection on the $\lambda$-factor of $\hat\partial^0_\infty(M)$, but is plainly consistent with the limiting connection of $\hat\nabla_\Lambda\Lambda = 0$ (while ignoring the geometry of the $\Sph2$ factor).

\subhead Null line at infinity
\endsubhead

Next, let's consider an actual absence of the $\Sph2$ factor: $\fcb(M)$ which is a null line instead of a null cone on $K$, i.e., the Cosmological Condition (3) is altered to $\int_{\rho_0}^\omega \frac1{a(\rho)^2} \,d\rho = \infty$.  This change in integral condition has no effect of any of the calculations for section 1; the only effect is on the topology of $\hat\partial^0_\omega(M)$, which is now just a line (null for $\omega = \infty$), parametrized by $\lambda \in \R1$; $K$ has disappeared from consideration.  Thus, no matter what choice is made for the transverse vector field in Theorem 2.1 (or the conformal factor in Theorem 2.2), we end up with $\hat\nabla_\Lambda\Lambda = 0$, and the connection is complete.  In the previous section, it was only the consideration of curves in $\Cal I^+$ which spin around the $K$ factor which necessitated close examination of acceleration in the $\Lambda$-component.  

But is this limit connection for $\fcb^0_\infty(M)$ the actual connection we want?  We should examine things more closely.

First note that for differentiability on $\fcb^0_\omega(M)$ to make sense as an extension of differentiability on $M$, we must retreat back to using only a sphere instead of a more general compact manifold $K$:  In order to achieve a manifold by adding (for the $\omega < \infty$ case) a line at $\{\omega\}$ to $\R1 \times (\alpha,\omega) \times K$, we must have $K$ be a sphere, for otherwise we don't have a local Euclidean topology when adding a point at $\{\omega\}$ to $(\alpha,\omega) \times K$; and similarly for the $\omega = \infty$ case, with adding a point at $\sigma = {\frac\pi2}$.  (If we have neither a manifold nor a manifold-with-boundary structure for $\hat M^0_\omega$, then there is no way of obtaining a differentiable structure on $\fcb^0_\omega(M)$ which is in any sense compatible with the differentiable structure on $M$.)

So what we have for $\hat M^0_\infty$ is not a manifold with boundary:  Rather, we have a pure manifold, exactly analogous to starting with $M$ being, say, $\R3$ with the $z$-axis removed, then adding the $z$-axis back for $\hat M$.  The differentiable structure is clear-cut, then, and the same co\"ordinates and co\"ordinate vector fields we employed in the previous section are applicable here: $\Lambda$, $\Sigma$, and $\Sph2$ vector fields.  And we might stop to ask about the extension of the spacetime metric or the spacetime connection to $\hat M^0_\infty$.  

The non-zero metric values are
$$\align
\<\Lambda,\Sigma\> & = -\frac12F(\rho)\sec^2(\sigma) \\
\<U,V\> & = F(\rho)a(\rho)^2h_{\Sph2}(U,V)
\endalign$$
with $\rho = \frac12(-\lambda + \tan\sigma)$.  If this extends to $\sigma = {\frac\pi2}$, that means we must have finite limits $p = \lim_{\rho \to \infty} -\frac12F(\rho)((2\rho + \lambda)^2 + 1)$ and $q = \lim_{\rho \to \infty}F(\rho)a(\rho)^2$ (as before, since we're really taking a limit as $\sigma \to {\frac\pi2}$, neither $p$ nor $q$ can have any $\lambda$-dependence; they must be constants).  Note that we must have $q \neq 0$ to get a non-degenerate metric.  Then for $\rho$ sufficiently large, we must have  $F(\rho)a(\rho)^2 > \frac12q$, or $\frac1{a(\rho)^2} < \frac2qF(\rho)$.  But we also have $F(\rho)((2\rho + \lambda)^2 + 1) < 2|p| + 1$ for $\rho$ sufficiently large,  or $F(\rho) < \frac{2|p| + 1}{(2\rho + \lambda)^2 + 1}$, yielding $\frac1{a(\rho)^2} < \frac2q(2|p| + 1)\frac1{(2\rho + \lambda)^2 + 1}$.  And that is incompatible with $\int_{\rho_0}^\infty \frac1{a(\rho)^2} d\rho = \infty$, our integral condition for having $\fcb^0_\infty(M)$ be a line.  It follows that the spacetime metric cannot extend to the boundary.

What
 about the spacetime connection?  We don't have a canonical simple spacetime to examine, so let's look at the general case.  The connection has values
$$\align
\nabla_\Lambda\Lambda & = -\frac12\frac{F'(\rho)}{F(\rho)}\Lambda \\
\nabla_\Lambda\Sigma & = 0 \\
\nabla_\Lambda V & = -\frac14\(\frac{F'(\rho)}{F(\rho)} + 2\frac{a'(\rho)}{a(\rho)}\)V \\
\nabla_\Sigma\Sigma & = \(\frac{F'(\rho)}{F(\rho)}\sec^2\sigma + 2\tan\sigma\)\Lambda \\
\nabla_\Sigma V & = \frac14\(\frac{F'(\rho)}{F(\rho)} + 2\frac{a'(\rho)}{a(\rho)}\)(\sec^2\sigma)V \\
\nabla_UV & = \frac12a(\rho)^2\(\frac{F'(\rho)}{F(\rho)} + 2\frac{a'(\rho)}{a(\rho)}\)h_{\Sph2}(U,V)(-(\cos^2\sigma)\Sigma + \Lambda) + \nabla^{\Sph2}_UV
\endalign$$
The Cosmological Conditions ($\frac{F'}F \to 0$, $\frac{a'}a \to 0$) give us that $\nabla_\Lambda\Lambda$ and $\nabla_\Lambda V$ both have limit 0 at $\fcb^0_\infty(M)$; but for $\nabla_\Sigma\Sigma$, $\nabla_\Sigma V$ and $\nabla_U V$ to have limits requires further specific behavior for $F(\rho)$ and $a(\rho)$: some spacetimes will manifest limits for the connection, while others won't.

So if we just want a connection $\hat\nabla$ on $\fcb^0_\infty(M)$, the natural choice is $\hat\nabla_\Lambda\Lambda = 0$; and this is complete.

(If we insist on looking at a null line at infinity for a compact factor $K \neq \Sph2$, then we must consider $\fcb^0_\infty(M)$ just to be a separate manifold in its own right, not differentiably associated with $M$, even though it is topologically embedded in $\hat M^0_\infty$.  In such a case the flow of the vector field $\Lambda$ on $M$ extends as a flow on $\fcb^0_\infty(M)$, so it makes sense to single out the vector field $\Lambda$ on that line; then the natural connection to take is, again, $\hat\nabla_\Lambda\Lambda = 0$, complete.)

\subhead Timelike boundary
\endsubhead

Next let us consider a timelike component of the causal boundary.  That means for, say, boundary at $\omega$, that $\omega$ is finite.  So this implies $\int_{\rho_0}^\omega \frac1{a(\rho)^2} d\rho < \infty$ for $a(\rho)$ increasing as $\rho \to \omega$, giving $B_\omega = K$, i.e., $\fcb^0_\omega(M) \cong \R1 \times K$.  The only way to obtain $B_\omega = *$ (i.e., $\fcb^0_\omega(M) \cong \R1$) is to employ the alternate condition of $a(\rho) \to 0$ eventually monotonically as $\rho \to \omega$ (or, similarly, at the $\alpha$-end of the range for $\rho$).

\subsubhead Timelike infinity
\endsubsubhead

For a boundary at infinity, the simplest example may be $\AdS$, Universal Anti-de Sitter space, i.e., the universal cover of Anti-de Sitter space (as in \cite{HwE}).  We can consider this space (with center removed) as the manifold $\R1 \times (0,\infty) \times \Sph2$ with metric 
$$ds^2 = g = -(\cosh^2 r)dt^2 + dr^2 + (\sinh^2 r)h_{\Sph2}$$
or, with $\rho = \tan^{-1}(\sinh r)$, 
$$g = (\sec^2\rho)(-dt^2 + d\rho^2 + (\sin^2\rho)h_{\Sph2})$$
with range of $\rho$ being $(\alpha,\omega) = (0, \frac\pi2)$.   (Note that we obtain the full spacetime by extending the range of $r$ from $(0,\infty)$ to $[0,\infty)$---or the range of $\rho$ from $(0,\frac\pi2)$ to $[0,\frac\pi2)$.  The addition of $\R1 \times \{0\} \times \Sph2$ still results in a static spherically symmetric spacetime, just not in the mode of using an open interval for the radial co\"ordinate.)  We have $\int_{\rho_0}^{\frac\pi2} \frac1{\sin^2(\rho)}\,d\rho = \cot(\rho_0) < \infty$, so $B_{\frac\pi2} = \Sph2$ (automatic, as noted above).  With $\omega$ finite, the natural co\"ordinates to use are $t$ and $\rho$; that is to say, $\widehat{\Min2_{0,\frac\pi2}}$ has the structure of a manifold with boundary  parametrized with $(t,\rho) \in \R1 \times (0,\frac\pi2]$ with boundary at $\rho = \frac\pi2$; by Theorem 1.1, then we have the same differentiable structure for $\AdS$, boundary at $\rho = \frac\pi2$.

So with a boundary of timelike nature, does the spacetime metric fare any better than in Minkowski space for extension to the boundary?  Let's examine the metric on the natural co\"ordinate vector fields $T = \frac\partial{\partial t}$, $P = \frac\partial{\partial\rho}$, and fields $U, V$ on $\Sph2$.  The non-zero elements are
$$\align
\<T,T\> & = -\sec^2\rho \\
\<P,P\> & = \sec^2\rho \\
\<U,V\> & = (\tan^2\rho)h_{\Sph2}(U,V)
\endalign$$
Clearly this does not extend to $\fcb^0_{\frac\pi2}(\AdS)$, i.e., to $\rho = \frac\pi2$.  So what about the connection?  We obtain
$$\align
\nabla_TT & = (\tan\rho)P \\
\nabla_TV & = 0 \\
\nabla_UV & = (\tan\rho)h_{\Sph2}(U,V)P + \nabla^{\Sph2}_UV \\
\nabla_TP & = (\tan\rho)T \\
\nabla_PP & = (\tan\rho)P \\
\nabla_PV & = \frac1{(\cos\rho)(\sin \rho)}V \\
\endalign$$
and we see that almost none of these extend to the boundary.  But for a connection on $\{\rho = \text{constant}\}$, we need look only at $T$ and the $\Sph2$-fields, the upper three lines.  Taking a cue from what worked in the null boundary case, let's look at surfaces $B^{\rho_0} = \{\rho = \rho_0\}$.  We're in much better shape with these than in the null case, as these surfaces are timelike:  That means we can ask for the induced metric connection, which we obtain just by projecting perpendicular to $P$.  This yields the connection $\nabla^{\rho_0}$ in $B^{\rho_0}$:
$$\align
\nabla^{\rho_0}_TT & = 0 \\
\nabla^{\rho_0}_TV & = 0 \\
\nabla^{\rho_0}_UV & = \nabla^{\Sph2}_UV 
\endalign$$
which has the obvious extension $\hat\nabla$ to $\fcb^0_{\frac\pi2}(\AdS)$:
$$\align
\hat\nabla_TT & = 0 \\
\hat\nabla_TV & = 0 \\
\hat\nabla_UV & = \nabla^{\Sph2}_UV 
\endalign$$
In other words, timelike infinity on $\AdS$ has, as the clearly natural connection, the product connection on $\R1 \times \Sph2$---and this is complete.

This result on $\AdS$ is perfectly general:

\proclaim{Theorem 3.1}
Let $M$ be a standard static spacetime of the form $M = \R1 \times (\alpha,\omega) \times K$ for $K$ a compact manifold, $\omega < \infty$, with $M$ having metric $ds^2 = F(\rho)(-dt^2 + d\rho^2 + a(\rho)^2 h_K)$ (where $h_K$ is a Riemannian metric on $K$), satisfying $a(\rho)$ is increasing on some interval $[\rho_+, \omega)$.  Then $\hat M^0_\omega$ has the structure $\R1 \times (\alpha,\omega] \times K$ with $\fcb^0_\omega(M)$ being $\R1 \times \{\omega\} \times K$ within that, a timelike hypersurface.  If $\nabla^{\rho_0}$ is the induced (metric) connection on each hypersurface $\{\rho = \rho_0\}$, then there is a connection $\hat\nabla = \lim_{\rho_0 \to \omega} \nabla^{\rho_0}$ on $\fcb^0_\omega(M)$; and that connection is the product connection $\R1 \times K$, which is complete.
\endproclaim

\demo{Proof}
We just need to examine the connection in $M$:
$$\align
\nabla_TT & = -\frac{F'(\rho)}{2F(\rho)}P \\
\nabla_TV & = 0 \\
\nabla_UV & = -a(\rho)^2\(\frac{F'(\rho)}{2F(\rho)} + \frac{a'(\rho)}{a(\rho)}\)h_K(U,V)P + \nabla^K_UV \\
\nabla_TP & = \frac{F'(\rho)}{2F(\rho)}T \\
\nabla_PP & = \frac{F'(\rho)}{2F(\rho)}P \\
\nabla_PV & = \(\frac{F'(\rho)}{2F(\rho)} + \frac{a'(\rho)}{a(\rho)}\)V \\
\endalign$$
On any (timelike) hypersurface $B^{\rho_0} = \{\rho = \rho_0\}$ we have the induced metric connection $\nabla^{\rho_0}$:
$$\align
\nabla^{\rho_0}_TT & = 0 \\
\nabla^{\rho_0}_TV & = 0 \\
\nabla^{\rho_0}_UV & = \nabla^K_UV 
\endalign$$
which has the obvious limit, as $\rho_0 \to \frac\pi2$, of $\hat\nabla$ on $\fcb^0_{\frac\pi2}(M)$:
$$\align
\hat\nabla_TT & = 0 \\
\hat\nabla_TV & = 0 \\
\hat\nabla_UV & = \nabla^K_UV 
\endalign$$
\qed 
\enddemo

If we employ a conformal model of the boundary, using a conformal factor which depends only on $\rho$, we get the exact same result:  The Levi-Civita connection for the conformal metric, projected to the $\{\rho = \omega\}$ hypersurface, is the product connection on $\R1 \times K$.  The conformal factor $G(\rho)$ needs to satisfy $G(\rho)F(\rho)$ and $G(\rho)F(\rho)a(\rho)^2$ both have positive, finite limits as $\rho \to \omega$, so there are some spacetimes of this form that do not have conformal boundaries, such as with $a(\rho) \to \infty$ as $\rho \to \omega$ but $F(\rho)$ stays bounded.

\subsubhead Timelike line at center or singularity
\endsubsubhead

$\AdS$ also provides a typical example of a ``boundary" being nothing more than a timelike line removed at the center of a spherically symmetric spacetime, simply for the convenience of using spherical symmetry in the co\"ordinate expression for the metric.  Recall that the manifold is given as $\R1 \times (0,\frac\pi2) \times \Sph2$ with $a(\rho) = \sin\rho$.  We have $a(\rho) \to 0$ monotonically as $\rho \to 0$ and $\int_0^{\rho_0} \frac1{a(\rho)^2} \,d\rho = \infty$, so $\fcb^0_0(\AdS)$ is a timelike line, added to $\AdS$ as $\{\rho = 0\}$ is added as a line in $\Min2_{0,\frac\pi2} \times \Sph2$.  In fact, the inclusion of this timelike line is not merely as a differentiable manifold, but it is fully metric, i.e., the metric on $\AdS$ extends smoothly to the additional line, making the question of connection utterly trivial---and, in fact, the line is complete in the metric connection.  This is set out in Theorem 3.2.

We can also have a timelike line for a singularity; the prototypical example is the singularity at $r = 0$ in  Reissner-Nordstr\"om, as explicated, for instance, in \cite{HwE}; its exposition in terms of causal boundary is found in \cite{FHr}.  Let $\Bbb R\Bbb N_{\text{int}}$ denote Interior Reissner-Nordstr\"om (i.e., behind the inner event horizon in the undercharged, $|q| < m$, case; behind the only event horizon in the critically charged case; and the entire spacetime in the overcharged case).  Then $\Bbb R\Bbb N_{\text{int}}$ follows the Common Model Spacetime Paradigm with $f(r) = 1 - \frac{2m}r + \frac{q^2}{r^2}$ with $r$ having range $(0,r_-)$ in the undercharged case ($r_- = m - \sqrt{m^2 - q^2}$).  The non-zero elements of the metric, in terms of co\"ordinate vectors $T = \frac\partial{\partial t}$, $R = \frac\partial{\partial r}$, and $U$ and $V$ in $\Sph2$, are
$$\align
\<T,T\>
 & = -f(r) \\
\<R,R\> & = \frac1{f(r)} \\
\<U,V\> & = r^2h_{\Sph2}(U,V)
\endalign$$
which clearly does not extend to $r = 0$.  Thus we must use the same technique as in Theorem 3.1 to find a suitable connection on $\fcb^0_0(\Bbb R\Bbb N_{\text{int}})$, suitably restricted to just a line.

\proclaim{Theorem 3.2}
Let $M$ be a standard static spacetime of the form $M = \R1 \times (\alpha,\omega) \times \Sph2$ with $\alpha >  -\infty$ and $M$ having metric $g = F(\rho)(-dt^2 + d\rho^2 + a(\rho)^2 h_{\Sph2})$, satisfying $a(\rho)$ is increasing on some interval $(\alpha,\rho_0]$ with $a(\rho) \to 0$ as $\rho \to \alpha$.

A) Suppose 
\roster
\item $a(\rho)$ is smoothly extendible to $\rho = \alpha$ ($a(\alpha) = 0$ as specified above) with $a'(\alpha) = 1$ and $a''(\alpha) = 0$; and  
\item $F(\rho)$ is smoothly extendible to $\rho = \alpha$ with $F(\alpha) > 0$ and $F'(\alpha) = 0$.  
\endroster
Then $\hat M^0_\alpha$ just adds a timelike line to $M$ at $\rho = \alpha$; and the metric smoothly extends to the added line.  Thus, there is the standard metric connection on $\fcb^0_\alpha(M)$; and it is $\nabla_TT = 0$, making the added line complete.

B) Suppose either (1) or (2) above fail, but $\int_\alpha^{\rho_0} \frac1{a(\rho)^2}\,d\rho = \infty$.  Then again $\hat M^0_\alpha$ just adds a timelike line to $M$ at $\rho = \alpha$; the spacetime metric does not extend to $\fcb^0_\alpha(M)$, but the same procedure as in Theorem 3.1 provides a limiting connection $\hat\nabla$ on that boundary: $\hat\nabla_TT = 0$, a complete connection.
\endproclaim

\demo{Proof} A)  Note that (1) implies that $B_\alpha = *$, so $\fcb^0_\alpha(M)$ is a timelike line:  We must have $\int_\alpha^{\rho_0} \frac1{a(\rho)^2}\,d\rho = \infty$.  For suppose not, i.e., suppose for some finite $A$, $\int_\alpha^{\rho_0} \frac1{a(\rho)^2}\,d\rho = A$; then with $a(\rho)$ increasing on $(\alpha,\rho_0]$ we see that for all $\rho \in (\alpha, \rho_0]$, $(\rho - \alpha)\frac1{a(\rho)^2} \le A$, or $a(\rho) \ge A^{-\frac12}\sqrt{\rho - \alpha}$. This implies $\frac{a(\rho)}{\rho - \alpha}  \ge A^{-\frac12}\frac1{\sqrt{\rho - \alpha}}$, which blows up as $\rho \to \alpha$, violating the extendibility of $\frac{a(\rho)}{\rho - \alpha}$ to $\alpha$.

To examine the extendibility of $g$ to $\{\rho = \alpha\}$, let us write $g$ as $g = F(\rho)(-dt^2 + \bar g)$, with $\bar g$ the implied metric on $N = (\alpha,\omega) \times \Sph2$, $\bar g = d\rho^2 + a(\rho)^2h_{\Sph2}$; and define co\"ordinates on $(\alpha,\omega) \times \Sph2$ that will extend to the center point.  We'll embed $N$ into Euclidean space $\R3$ via $(\rho,p) \mapsto (\rho - \alpha)i(p)$ for $i: \Sph2 \to \R3$ the standard embedding of the unit sphere; then let $x^1, x^2, x^3$ be the standard rectangular co\"ordinates on $\R3$, transferred to $N$; and if $i(p) = (u^1, u^2, u^3)$, then $x^i(\rho, p) = (\rho - \alpha)u^i$.  

Now in $\R3$ we have the Euclidean metric $h_{\text{Euc}}$, which we can use to define the projections $\parallel$ and $\perp$ from $T_x\R3$ to, respectively, the subspaces of vectors parallel and perpendicular to the embedded 2-sphere around the origin in $\R3$ passing through $x$: $v^\perp = h_{\text{Euc}}(v,p)p$ (where $x = i(\rho,p)$) and $v^\parallel = v - v^\perp$.  We can use these projections for calculations with $\bar g$:
$$\align
\bar g(v,w) & = \bar g(v^\perp + v^\parallel, w^\perp + w^\parallel) \\
& = \bar g(v^\perp, w^\perp) + \bar g(v^\parallel, w^\parallel) \\
& = h_{\text{Euc}}(v,p)h_{\text{Euc}}(w,p)\bar g(p,p) + \bar g(v^\parallel, w^\parallel) \\
& = h_{\text{Euc}}(v,p)h_{\text{Euc}}(w,p) + a(\rho)^2h_{\Sph2}(v^\parallel, w^\parallel)
\endalign$$
Let $e_i = \frac\partial{\partial x^i}$; then $h_{\text{Euc}}(e_i,p) = u^i$, where $p$ has co\"ordinates $(u^1, u^2, u^3)$.  This gives us 
$$\align
e_1^\perp & = ((u^1)^2, u^1u^2, u^1u^3) \\
e_1^\parallel & = (1 - (u^1)^2, -u^1u^2, -u^1u^3)
\endalign$$
and so on. Note that $h_{\text{Euc}} = d\rho^2 + (\rho - \alpha)^2h_{\Sph2}$, so that, restricted to the $\parallel$-space, we have $h_{\Sph2} = \frac1{(\rho - \alpha)^2}h_{\text{Euc}}$.  This gives us
$$\align
\text{for all }i, \; \bar g(e_i,e_i) & = (u^i)^2 + \frac{a(\rho)^2}{(\rho - \alpha)^2}h_{\text{Euc}}(e_i^\parallel, e_i^\parallel) \\
& = (u^i)^2 + \frac{a(\rho)^2}{(\rho - \alpha)^2}\((1 - (u^i)^2)^2 + (u^i)^2\sum_{k \neq i}(u^k)^2\) \\
& = (u^i)^2 + \frac{a(\rho)^2}{(\rho - \alpha)^2}\(1 - 2(u^i)^2 + (u^i)^4 + (u^i)^2(1 - (u^i)^2)\) \\ 
& = (u^i)^2 + \frac{a(\rho)^2}{(\rho - \alpha)^2}\(1 - (u^i)^2\) \\
& = \frac{a(\rho)^2}{(\rho - \alpha)^2} + \(1 - \frac{a(\rho)^2}{(\rho - \alpha)^2}\)(u^i)^2
\endalign$$
and
$$\align
\text{for all }i \neq j, \; \bar g(e_i, e_j) & = u^iu^j + \frac{a(\rho)^2}{(\rho - \alpha)^2}h_{\text{Euc}}(e_i^\parallel, e_j^\parallel) \\
& = u^iu^j + \frac{a(\rho)^2}{(\rho - \alpha)^2}\((1 - (u^i)^2)(-u^iu^j) + (-u^iu^j)(1 - (u^j)^2) \right. \\ 
& \quad\quad\quad\quad\quad\quad\quad\quad\quad\quad \left. +\; u^iu^j(u^k)^2\) \quad(\text{where }k \neq i, j) \\
& = u^iu^j + \frac{a(\rho)^2}{(\rho - \alpha)^2}u^iu^j\(-1 - 1 + (u^i)^2 + (u^j)^2 + (u^k)^2\) \\
& = u^iu^j - \frac{a(\rho)^2}{(\rho - \alpha)^2}u^iu^j \\
& = \(1 - \frac{a(\rho)^2}{(\rho - \alpha)^2}\)u^iu^j
\endalign$$
yielding the expression for $\bar g$ in the $x^i$ co\"ordinates:
$$\bar g  = \(\frac{a(\rho)}{\rho - \alpha}\)^2\sum_i(dx^i)^2 + \(1 - \(\frac{a(\rho)}{\rho - \alpha}\)^2\)\sum_{i,j}u^iu^j\,dx^idx^j $$
and thus for $g$:
$$g = F(\rho)\(-dt^2 + \(\frac{a(\rho)}{\rho - \alpha}\)^2\sum_i(dx^i)^2 +\(1 - \(\frac{a(\rho)}{\rho - \alpha}\)^2\)\sum_{i,j}u^iu^j\,dx^idx^j \)$$
(We could express it in terms of $x^i$ instead of $u^i$ and $\rho$, but there is advantage to doing it this way.)

What is necessary for this metric to extend smoothly to $\rho = \alpha$?  Clearly we need $F(\alpha) > 0$.  Considering just the last sum, we must have $a'(\alpha) = 1$, since that term must vanish at $\rho = \alpha$, as $u^i$ is not defined there.  Now fix a point $p \in \Sph2$ and consider the curve in $\R3$ $\sigma$ given by $\sigma(s) = sp$.  This corresponds to a curve in $M$ (for fixed $t$), with $\rho = \alpha + |s|$, and $g$ is obliged to be smooth along this curve.  In particular $g(T,T) = -F(\alpha + |s|)$, and the only way this can be smooth at $s = 0$ is with $F'(\alpha) = 0$.  Similarly, we need $F(\rho)\(\frac{a(\rho)}{\rho - \alpha}\)^2$ to have a vanishing derivative at $\rho = \alpha$; use of l'H\^opital shows this comes down to $a''(\alpha) = 0$.  But with all these conditions met, $g$ is smoothly extendible to $\rho = \alpha$.

B) The paragraph above shows that failure of conditions (1) or (2) implies the metric does not extend to $\rho = \alpha$.  But the procedure in Theorem 3.1 provides a limiting connection $\hat\nabla$ on $\fcb^0_\alpha(M)$, yielding $\hat\nabla_TT = 0$, a complete connection.
\qed

\enddemo

\subhead Spacelike singularity
\endsubhead 

To complete the picture of boundaries for standard-static, spherically symmetric spactimes, let's look at the generalization afforded by spactimes which experience a phase transition for the Killing field from timelike to spacelike, most iconically realized with Schwarzschild space, $\Sch$.  Recall (\cite{HwE}) that extended Schwarzschild can be expressed as $\R1 \times (0, \infty) \times \Sph2$ with metric $g = -(1 - \frac{2m}r)dt^2 + (1 - \frac{2m}r)^{-1}dr^2 + r^2 h_{\Sph2}$, with the killing field $T = \frac{\partial}{\partial t}$ timelike in $\Sch\ext$ ($r > 2m$) and spacelike in $\Sch\Int$ ($r < 2m$); the transition at $r = 2m$ is the event horizon, and the singularity occurs at $r = 0$---not part of the spacetime, but it is $\fcb(\Sch\Int)$.

We briefly note that the metric does not extend to the singularity:  With $R = \frac{\partial}{\partial r}$ and $U, V$ co\"ordinate vectors in $\Sph 2$ we have
$$\align
\<R,R\> & = -(\frac{2m}r - 1)^{-1} \\
\<T,T\> & = \frac{2m}r - 1 \\
\<U,V\> & = r^2h_{\Sph2}(U,V)
\endalign$$
which does all the wrong things as $r \to 0$.  But we can handle this in essentially the same was as in Theorem 3.1.  As shown in \cite{Hr2}, $\fcb(\Sch\Int)$ is a spacelike $\R1\times\Sph2$ occupying ${r = 0}$ in $\widehat{\Sch\Int} \cong \R1\times[0,2m)\times\Sph2$; and the $r = r_0$ hypersurfaces have induced connections which converge to the product connection at $r = 0$.  And we can do this quite generally.  

It is more convenient, for these spactimes exhibiting phase transition for the Killing field, to express the metric as we've just done for $\Sch$, using $r$ instead of $\rho$; we'll label the portion where the Killing field is timelike, the exterior part of the spacetime, and the portion where the Killing field is spacelike, the interior part of the spacetime. 

\proclaim{Theorem 3.3} Let $M = \R1 \times (\rmin,\rmax) \times K$, $K$ compact, 
with metric expressible on most of $M$ as $g = -f_1(r)dt^2 + f_2(r)dr^2 + b(r)^2h_K$ for $h_K$ a Riemannian metric on $K$ and $f_1$, $f_2$, and $b$ functions on $(\rmin,\rmax)$ such that for some $\reh \in (\rmin,\rmax)$,
\roster
\item $f_1 > 0$ and $f_2 > 0$ on $(\reh,\rmax)$ 
\item $f_1 < 0$ and $f_2 < 0$ on $(\rmin,\reh)$ 
\item $b > 0$ on $(\rmin,\rmax)$.
\endroster
Let $M\Int = \R1 \times (\rmin,\reh) \times K$ and $M\ext \cong \R1 \times (\reh,\rmax) \times K$; these two portions are where the expression above for $g$ is valid.

Suppose that for some $r_1 \in (\rmin,\reh)$, $\int_{\rmin}^{r_1} \sqrt{\frac{f_2(r)}{f_1(r)}} \,dr < \infty$.  

A) If $\int_{\rmin}^{r_1} \frac{\sqrt{-f_2(r)}}{b(r)}\,dr < \infty$ also,  then $\widehat{M\Int} \cong \R1 \times [\rmin,\reh) \times K$ and $\fcb(M\Int) \cong \R1 \times K$, spacelike, fitting into $\widehat{M\Int}$ as $\R1 \times \{\rmin\} \times K$; this yields the differentiable structure on $\fcb(M\Int)$.  The same procedure as in Theorem 3.1---letting $\nabla^{r_0}$ denote the induced connection
 on $\{r = r_0\}$ and letting $\hat\nabla$ be the limit of $\nabla^{r_0}$ as $r_0 \to \rmin$---yields a complete connection $\hat\nabla$ on $\fcb(M\Int)$: the product connection on $\R1 \times K$.

B) If $\int_{\rmin}^{r_1} \frac{\sqrt{-f_2(r)}}{b(r)}\,dr = \infty$, then the only topological/causal difference from that above is that the $K$-factor in $\fcb(M\Int)$ is reduced to a point, i.e., $\fcb(M\Int) \cong \R1$.  If, in addition, $K = \Sph2$, then we have a differentiable structure for $\fcb(M\Int)$, and the same limit procedure with $\nabla^{r_0}$ results in $\hat\nabla$ being the standard connection in $\R1$: complete.
\endproclaim

\demo{Proof} First let us note that in $M\Int$ (where $R = \frac{\partial}{\partial r}$ is timelike) it is the $r = \rmin$ end of $(\rmin,\reh)$ which is the future:  In $M\ext$ we take, as usual, $T = \frac{\partial}{\partial t}$ to be future-directed. Then the curve $\beta(s) = (t(s),r(s),p)$, for fixed $p \in K$, is a future-directed null curve headed from $M\ext$ towards the event horizon $M\eh = \R1 \times \{\reh\}\times K$ so long as  $f_1(r(s))t'(s)^2 = f_2(r(s))r'(s)^2$, $t'(s) > 0$, and $r'(s) < 0$; and thus $\beta$ will remain a future-directed null curve, across $M\eh$ and into $M\Int$, with the same conditions.  Thus, it is $-R$ that is future-timelike in $M\Int$.  

Concentrating on $M\Int$, let us write the metric as 
$$g = -f_2(r)\(-dr^2 + \frac{f_1(r)}{f_2(r)}dt^2 + \frac{b(r)^2}{-f_2(r)}h_K\)$$

Proposition 3.5 of \cite{Hr2} addresses the future causal boundary of a multiply warped product spacetime of the form $N = (a,b) \times N_1 \times \cdots \times N_m$ with metric $-dr^2 + \sum_i \phi_i h_i$ with each $h_i$ a  Riemannian metric on the manifold $N_i$ and $\phi_i: (a,b) \to (0,\infty)$.  By 3.5(c), if each $h_i$ is a complete metric and (for some $c \in (a,b)$) $I_i = \int_a^c \phi_i^{-1} = \infty$ occurs only for $K_i$ compact, then (taking $b$ to be the future end of the interval), $\fcb(N)$ has the topology of the product $\prod_{I_i < \infty} N_i$, fitting into $\hat N$ as $(a,b] \times \prod_{i =1}^m N_i$ by identifying $\{b\} \times \prod_{I_i = \infty} N_i$ to a single point.  We have $M\Int$ conformal to such a spacetime with $(a,b) = (\rmin,\reh)$, $(N_1, h_1) = (\R1, dt^2)$, $(N_2, h_2) = (\Sph2, h_{\Sph2})$, $\phi_1 = \frac{f_1}{f_2}$, and $\phi_2 = \frac{b^2}{-f_2}$.  By assumption we have $\int_{\rmin}^{r_1} \phi_1^{-\frac12} < \infty$, and we have $N_2$ compact.  Then the results follow, in both (A) and (B), concerning the topological, differentiable, and causal characteristics of $\fcb(M\Int)$.

To look at the connection, we'll use  $T$ and $R$ as above with $U, V$ co\"ordinate vectors in $K$.  We then get
$$\align
\nabla_RR & = \frac{f_2'(r)}{2f_2(r)}R \\
\nabla_RT & = \frac{f_1'(r)}{2f_1(r)}T \\
\nabla_RV & = \frac{b'(r)}{b(r)}V \\
\nabla_TT & = \frac{f_1'(r)}{2f_2(r)}R \\
\nabla_TV & = 0 \\
\nabla_UV & = -\frac{b(r)b'(r)}{f_2(r)}h_K(U,V)R + \nabla^K_UV
\endalign$$
Then the induced connection $\nabla^{r_0}$ on $\{r = r_0\}$ is the product connection on $\R1 \times K$.  In case (A), so is the limit $\hat\nabla = \lim_{r_0 \to \rmin} \nabla^{r_0}$.  In case (B), we just note we have the limit $\hat\nabla_TT = 0$. \qed
  
\enddemo

\remark{3.4 Remark on extending across event horizons}In Theorem 3.3, we explicitly assumed that we had a spactime encompassing the interior and the exterior sections, as well as the event horizon separating them.  What if we just start with $M\ext$ and functions $f_1$ and $f_2$ behaving as required on $(\rmin,\rmax)$---do we necessarily obtain a spacetime that extends from $M\ext$ across $M\eh$ and including $M\Int$?  We can get a partial answer from examining causal boundaries:  By expressing $g$ as 
$$g = f_1(r)\(-dt^2 + \frac{f_2(r)}{f_1(r)}dr^2 + \frac{b(r)^2}{f_1(r)}h_K\)$$ for examination of $M\ext$ by the methods already used in section 2 here, and by using
the discussion in section 6.1.1 of \cite{FHr} for examination of the past causal boundary of $M\Int$ utilizing the conformal expression for $g$ in Theorem 3.3, we obtain these results:  

Let $(\kappa,\omega)$ for $\rho$ correspond to $(\reh,\rmax)$ for $r$, where $d\rho = \sqrt{\frac{f_2(r)}{f_1(r)}}\,dr$.  Assume that for some $r_-, r_+ \in (\reh,\rmax)$, $\frac{b^2}{f_1}$ is decreasing on $(\reh,r_-)$ and increasing on $(r_+,\rmax)$.  Then
\roster
\item $\fcb^0_\omega(M\ext)$ is null if $\int_{r_+}^{\rmax} \sqrt{\frac{f_2}{f_1}} = \infty$, otherwise timelike.  
\item $\fcb^0_\omega(M\ext) \cong \R1 \times K$ if $\int_{r_+}^{\rmax} \frac{\sqrt{f_1f_2}}{b^2} < \infty$, otherwise it is $\R1$.
\item $\fcb^0_\kappa(M\ext)$ is null if $\int_{\reh}^{r_-} \sqrt{\frac{f_2}{f_1}} = \infty$, otherwise timelike.  
\item $\fcb^0_\kappa(M\ext) \cong \R1 \times K$ if $\int_{\reh}^{r_-} \frac{\sqrt{f_1f_2}}{b^2} < \infty$, otherwise it is $\R1$.
\item If (for some $r_0 \in (\rmin,\reh)$), $\int_{r_0}^{\reh} \sqrt{\frac{f_2}{f_1}} = \infty$, then $\check\partial(M\Int)$ consists of two null cylinders on $K$, each terminating in the past on the same spacelike copy of $K$; in effect, $\check\partial^0_\infty(M\Int)$ and $\check\partial^0_{-\infty}(M\Int)$ both consist of $\R1\times K$, null in the $\R1$ factor, where superscript 0 denotes ignoring the extended spacelike $i^-$ (a copy of $K$), and the subscripts indicate the $t = \infty$ or $t = -\infty$ ends of the past causal boundary.
\item If $\int_{r_0}^{\reh} \sqrt{\frac{f_2}{f_1}} < \infty$, then $\check\partial(M\Int)$ is spacelike; if $\int_{r_0}^{\reh} \frac{\sqrt{-f_2}}b < \infty$, then $\check\partial(M\Int) \cong \R1 \times K$, and otherwise it is $\R1$.
\item If $\int_{\rmin}^{r_0} \sqrt{\frac{f_2}{f_1}} = \infty$, then $\fcb(M\Int)$ consists of two null cylinders on $K$, each terminating in the future on the same spacelike copy of $K$ (spacelike extended $i^+$ a copy of $K$, $\fcb^0_\infty(M\Int)$ and $\fcb^0_{-\infty}(M\Int)$ both $\R1\times K$).
\item If $\int_{\rmin}^{r_0} \sqrt{\frac{f_2}{f_1}} < \infty$, then $\fcb(M\Int)$ is spacelike; if $\int_{\rmin}^{r_0} \frac{\sqrt{-f_2}}b < \infty$, then $\fcb(M\Int) \cong \R1 \times K$, and otherwise it is $\R1$.
\endroster

The classic event horizon, as in $\Sch$ or $\Bbb R\Bbb N$, is permitted by the topological/causal information when the  integrals in (3) and (5) are infinite, and that in (4) is finite, as that permits marriage of a null $\R1 \times K$ in the future causal boundary of $M\ext$ with a null $\R1 \times K$ in the past causal boundary of $M\Int$ (both at $r = \reh$).

\endremark

\enddocument